\documentclass[pre,showpacs,nofootinbib,twocolumn,superscriptaddress]{revtex4}
\usepackage[latin1]{inputenc}
\usepackage{graphicx,psfrag}
\usepackage{bm,longtable}

\usepackage{txfonts}

\newcommand{\E}{\mathrm{e}}

\newcommand{\average}[1]{\left\langle{#1}\right\rangle}

\newcommand{\eij}{\epsilon_{ij}}
\newcommand{\dij}{\Delta_{ij}}
\newcommand{\sij}{\sigma_{ij}}
\newcommand{\lij}{l_{ij}}

\newcommand{\tu}{\tau_{\mathrm u}}
\newcommand{\p}[1]{\left({#1}\right)}
\newcommand{\pq}[1]{\left[{#1}\right]}
\newcommand{\pg}[1]{\left\{{#1}\right\}}

\newcommand{\Eu}{E_{\mathrm u}}
\newcommand{\xu}{x_{\mathrm u}}
\newcommand{\Lu}{L_{\mathrm u}}
\newcommand{\Ze}{{\mathcal Z}}

\begin{document}
\title{Protein mechanical unfolding: a model with binary variables}
\author{A. Imparato}
\email{alberto.imparato@polito.it}
\affiliation{Dipartimento di Fisica and CNISM, Politecnico di Torino,
  c. Duca degli Abruzzi 24, Torino, Italy}
\affiliation{INFN, Sezione di Torino, Torino, Italy}
\author{A. Pelizzola}
\email{alessandro.pelizzola@polito.it}
\affiliation{Dipartimento di Fisica and CNISM, Politecnico di Torino,
  c. Duca degli Abruzzi 24, Torino, Italy}
\affiliation{INFN, Sezione di Torino, Torino, Italy}
\author{M. Zamparo}
\email{marco.zamparo@polito.it}
\affiliation{Dipartimento di Fisica and CNISM, Politecnico di Torino,
  c. Duca degli Abruzzi 24, Torino, Italy}

\begin{abstract}
A simple lattice model, recently introduced as a generalization of the
Wako--Sait\^o model of protein folding, is used to investigate the
properties of widely studied molecules under external forces.  The
equilibrium properties of the model proteins, together with their
energy landscape, are studied on the basis of the exact solution of
the model.  Afterwards, the kinetic response of the molecules to a force is
considered, discussing both force clamp and dynamic loading protocols
and showing that theoretical expectations are verified. The kinetic
parameters characterizing the protein unfolding are evaluated by using
computer simulations and agree nicely with experimental results, when
these are available. Finally, the extended Jarzynski equality is
exploited to investigate the possibility of reconstructing the free
energy landscape of proteins with pulling experiments.

\end{abstract}
\maketitle

\section{Introduction}

The three-dimensional structure of proteins is strictly connected to
the biological functions these molecules perform in living cells
\cite{ABL}. Among various experimental techniques, an increasingly
important role in the study of protein structures is being played by single--molecule force spectroscopy,
where  proteins \cite{KSGB,rgo,cv1,DBBR,Ober1,exp_fc1,exp_fc2} (but also  nucleic acid fragments \cite{NA_exp}) are pulled by applying  controlled forces to their  ends through an atomic force microscope (AFM) or optical tweezers. 

By studying the dynamical response of proteins to constant or varying
loading, much information on their structure has been gathered
\cite{KSGB,rgo,cv1,DBBR, Ober1}.  In particular, the possibility of controlling
the applied force with high precision has allowed to trace the
molecule folding and unfolding pathways \cite{DBBR,exp_fc1,exp_fc2}.
Nevertheless, as the size of the molecules increases, force
spectroscopy outcomes cannot be easily related to molecular
properties.  Thus theoretical models of biomolecules subject to an
external force have been developed \cite{thiru,ciep,LKH,ILT},
which represent important tools to study the interplay between protein
response to external force and molecular structure.

In most cases, these models are based on a coarse--grained description
of the biomolecule, their dynamical degrees of freedom being related
to the coordinates and velocities of a suitable set of reference
beads, typically one or a few per amino~acid \cite{thiru,ciep,LKH,ILT}. Equilibrium and
nonequilibrium results are then obtained by means of (time 
expensive) computer simulations. 

In a recent paper \cite{IPZ} we have approached the problem of protein unzipping from a different point of view,
introducing a simple lattice model with binary degrees of freedom,
based on (and generalizing) the Wako--Sait\^o (WS) model of protein
folding \cite{WS1,WS2}. Despite its simplicity, the model turned
out to exhibit the typical response of real proteins to pulling. In
particular, the mechanical unfolding of the 27th immunoglobulin domain
of titin was investigated, considering both force clamp and dynamic
loading protocols. Theoretically expected laws were verified and an
excellent agreement with the experimental values of the characteristic
kinetic parameter was found. The model was also used to investigate
the possibility of reconstructing the protein free energy landscape by
exploiting an extended version of the Jarzynski equality (JE)
\cite{jarz,HumSza,alb1}.

The present paper has several purposes.
On the methodological
side, we give a full derivation of the relation between our model and
the original WS model, and show in detail how to obtain free energy
landscapes. On the application side, we consider three other molecules
(including BBL, whose thermal behaviour has recently been the subject
of some debate, see ref.~\cite{SFM} and references therein, and ref.~\cite{dib}) in addition to titin and, together with the properties
already investigated in \cite{IPZ}, we discuss also the probability
distribution of unfolding times and forces.

The article is organized as follows. In Section \ref{model}, we
describe the details of our model, its connection with the WS model,
and the simulation method. In Section \ref{eq_p} we present the
equilibrium properties of the model proteins, as functions of force
and temperature. We then present the results of mechanical unfolding
obtained by numerical simulations, first, in Section \ref{FC}, for the force clamp manipulation  and then, in Section \ref{DL}, for the dynamic loading. In Section \ref{EJE}, the free energy landscape of
the model proteins is reconstructed from unfolding manipulations, by
using an extended Jarzynski Equality. Conclusions are drawn, and future developments
are sketched, in Section \ref{Conclusions}.

\section{The model}\label{model}

In the present section, we define our model and show that in the absence of
an external force it reduces to the WS model of protein folding. The
latter was introduced in 1978 by Wako and Sait\^o, in two 
papers \cite{WS1,WS2} that appear to have been forgotten until recent
years. The same model was independently reintroduced by Mu\~noz, Eaton
and coworkers at the end of the '90s \cite{ME1,ME2,ME3}. These authors
used the model to describe and interpret experimental results, and
soon the model became quite popular \cite{Amos,CCBM,ItohSasai,AbeWako,TD1,Ap1,Ap2,ZP0,ZP,BPZ1,BPZ2}. This is why it is
often referred to as the Mu\~noz--Eaton, or
Wako--Sait\^o--Mu\~noz--Eaton, model. The first recent reference to
the original work by Wako and Sait\^o appeared, as far as we know, in
\cite{ItohSasai}. 

Similarly to the WS model, the state of a protein is defined
according to the conformation of its peptide bond backbone, and in
order to reduce the degrees of freedom,  we assume that the peptide bonds can exist
only in two conformations: native and non-native. Thus a $N+1$
amino~acid protein is represented as a chain of $N$ peptide bonds, and
a binary variable $m_{k}$ is associated to each peptide bond. Bonds
are numbered from 1 to $N$ and amino~acids from 1 to $N+1$, bond $k$
connecting amino~acids $k$ and $k+1$. The variable $m_{k}$ takes the
value $0,1$ corresponding to a peptide bond in unfolded or native
state respectively. 

In order to couple the molecule to an external force we regard
stretches of consecutive native bonds (delimited by unfolded bonds) as
rigid portions of the protein with their own (native) end-to-end
length: in the following $\lij$ will indicate the end-to-end length of
the stretch of consecutive amino~acids connected by native bonds and
delimited by unfolded bonds in positions $i$ and $j > i$: $m_i = m_j =
0$, $m_k = 1$ for $i < k < j$; if we define the quantity  
\begin{equation}
S_{ij}(m)\equiv
(1-m_{i})(1-m_j)\prod_{k=i+1}^{j-1} m_k, 
\end{equation} 
this condition can be expressed 
in a more compact form: $S_{ij}(m) = 1$.  In the limiting case $j - i = 1$ the stretch
reduces to amino~acid $j = i+1$, which is characterized by its native
length $l_{i,i+1}$. In the following, ``stretch'' will refer to
stretches of any length, including single amino~acids. Boundary
conditions $m_0 = m_{N+1} = 0$ are introduced to define the stretches
at the protein ends. It is easy to verify that the number of stretches
is equal to 1 plus the number of 0's in the set $\{m_k,\,  k = 1, \ldots
N\}$.

\begin{figure}[h]
\center
\psfrag{lij}[cb][cb][1.]{$\lij$}
\psfrag{L}[tt][tt][1.]{$L$}
\psfrag{f}[cb][cb][1.]{$f$}
\psfrag{zeta}[lt][lt][1.]{$\zeta$}
\includegraphics[width=8cm]{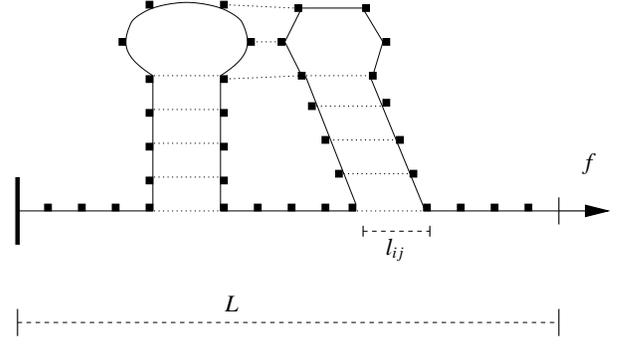}
\caption{Cartoon of the model protein, with a force applied to
one of the free ends. Dots denote amino~acids and dashed lines
denote contacts.}
\label{mol1}
\end{figure}

We want to keep our model as simple as possible, therefore only two
orientations of rigid stretches are
considered: given the direction of the external force, a stretch can
only be oriented parallel or antiparallel to the force direction.
Thus, following \cite{IPZ}, we introduce the variables $\sigma_{ij}=\pm 1$ which describe the orientation of the
stretch with respect to the external force $f$.  Therefore, the
configuration of the molecule is fixed by the set of
$(\{m_k\},\{\sij\})$ values, and for each configuration the protein
end--to--end length is given by
\begin{equation}
L(\pg{m_k}, \{ \sigma_{ij}\}) = \sum_{0 \le i < j \le N+1} \lij
\sigma_{ij} S_{ij}(m).
\label{L_def}
\end{equation}
A cartoon of the model protein is plotted in fig.\ref{mol1}. 
It is worth noting that a given bond configuration $\{m_k\}$ dynamically fixes the 
set of variables $\{\sij\}$: for each configuration $\{m_k\}$ only those variables $\{\sij\}$ such that $S_{ij}(m)=1$ are considered.

As in the original WS model \cite{WS1,WS2,ME1,ME2,ME3}, in the present model
two amino~acids $i$ and $j+1 > i$ interact only if {\it all} the
peptide bonds connecting them along the chain (that is, bonds from $i$
to $j$) are in the native state, and if they are close enough in the
native configuration.  The Hamiltonian of the model reads thus
\begin{equation}
\mathcal H(\pg{m_k},\{\sigma_{ij}\},f)=-\sum_{i=1}^{N-1} \sum_{j=i+1}^N \eij \dij
\prod_{k=i}^j m_k -f L(\pg{m_k}, \{ \sigma_{ij}\})
\label{hnoi}
\end{equation} 

The quantity $\eij>0$ represents the interaction energy between the
amino~acids $i$ and $j+1$, (defined as in \cite{ME3,Ap1,BPZ1}, see below
for details), while $\dij=0,1$ is the corresponding element of the
contact matrix, taking the value  1 if the distance between any
two atoms of the two amino~acids is smaller than some threshold distance, 
or the value 0
if none of all the possible atom pairs satisfies this condition, see discussion below.

The microscopic degrees of freedom $\sigma_{ij}$ do not interact among
themselves, hence the partial partition sum over them can be performed
analytically. We obtain the effective Hamiltonian $\mathcal H_{\rm
eff}$, defined by
\[
\sum_{\{\sigma_{ij}\}} \exp\pq{- \beta \mathcal
H(\pg{m_k},\{\sigma_{ij}\},f)} = \exp\pq{- \beta \mathcal H_{\rm
eff}(\pg{m_k},f)},
\]
that reads
\begin{eqnarray}
&& \mathcal H_{\rm eff}(\pg{m_k},f) = -\sum_{i=1}^{N-1} \sum_{j=i+1}^N
\eij \dij \prod_{k=i}^j m_k\nonumber \\
&& -  k_BT \sum_{0 \le i < j \le N+1} 
\ln \left[ 2 \, {\rm cosh} \left( \beta f l_{ij} \right) \right]
S_{ij}(m).
\label{heff}
\end{eqnarray}
This effective Hamiltonian is a linear combination of products of
consecutive $m_k$'s (including the single peptide bond) and therefore
it has the same mathematical structure of the WS model, whose
Hamiltonian reads
\begin{equation}
\mathcal H_0(\pg{m_k})=-\sum_{i=1}^{N-1} \sum_{j=i+1}^N \eij \dij
\prod_{k=i}^j m_k -k_B T \sum_{i=1}^N q_i(1-m_i),
\label{H0}
\end{equation} 
where $q_i$ is the entropic cost of ordering bond $i$.
Given this similarity, the partition function associated to the
Hamiltonian (\ref{heff}) can be summed exactly, and the thermodynamic
quantities can be obtained, as discussed in ref.~\cite{Ap1,Ap2}. 
In the case $f = 0$, $\mathcal H_{\rm eff}$ reduces to (up to an
additive constant)
\[
\mathcal H_{\rm eff}(\pg{m_k},0) = -\sum_{i=1}^{N-1} \sum_{j=i+1}^N
\eij \dij \prod_{k=i}^jm_k - k_B T \ln 2 \sum_{i=1}^N (1-m_i),
\]
which corresponds to the WS model, eq.(\ref{H0}), with $q_i = \ln 2$.
In the absence of force, our new orientational degrees of freedom
correspond to an entropic gain $q_i$ associated to the unfolded
peptide bonds. This quantity takes the value $\ln 2$ because we have made the
simplest choice of two possible orientations per stretch. Of course
this is a very rough discretization of the actual orientational
degrees of freedom of the main chain, namely the dihedral angles. In
principle, one could think to orientational variables taking values in
a larger set, and this would yield different, even non--uniform, $q_i$
values.

As discussed in ref.~\cite{IPZ}, in the present model the quantities
$\dij\, ,\eij$ and $\lij$ are chosen by analyzing the native structure
of a given molecule taken by the Protein Data Bank (pdb in the
following, http://www.pdb.org/).  We briefly review here the choice
criteria for readability's sake.  Two amino~acids $i$ and $j+1$ (with
$j+1>i+2$) are in contact ($\Delta_{ij}=1$) if, in the native state of
the protein, at least two atoms from these amino~acids are closer than
$4\, \AA$. In this case $\eij$ is taken to be equal to $k\epsilon$,
where $k$ is an integer such that $5(k-1)<n_{at}\leq 5k$, and $n_{at}$
is the number of atoms of the two amino~acids whose distance is not
larger than the threshold distance.  The quantity $\epsilon$ is the
protein energy scale, and is determined by imposing that, at zero
force and at the experimental denaturation temperature $T_m$, the
fraction of folded molecules is $p(T_m) = 1/2$. In order to estimate
$p$ we introduce the number of native peptide bonds $M = \sum_{k=1}^N
m_k$ and its average density $m = \langle M \rangle/N$, which in the
following will be used as an order parameter. The fraction of folded
molecules is then estimated, assuming a two--state picture, as $p =
(m - m_\infty)/(m_0 - m_\infty)$, where $m_\infty = 1/3$ is the value
of $m$ at infinite temperature, while $m_0$ is a good representative
of the folded state \cite{BPZ1}. For most molecules we can take $m_0 = m(T=0) =
1$, an exception being the WW domain of PIN1 (pdb code 1I6C), which orders perfectly
only at zero temperature and exhibits a wide plateau in m(T) in the
range 200--300 K \cite{BPZ1}. In this case we choose $m_0 =
m(T=273 K) < 1$.

As far as the parameters $\lij$ are concerned, the generic amino~acid
$i$ is represented by its $N_{i}-C_{\alpha,i}-C_{i}$ sequence.  Taking
the native state as the reference configuration, $\lij$ is chosen as
the distance between the midpoint of the $C_{i}$ and $N_{i+1}$ atoms
and the midpoint of the $C_{j}$ and $N_{j+1}$ atoms.

In the present paper we shall consider four different molecules of
increasing size: 1BBL (37 amino~acids, see \cite{BPZ1} for details), 1I6C (39 amino~acids), 1COA (64
amino~acids), 1TIT (89 amino~acids).  Here and in the following the
proteins are indicated with their pdb code. Some results on the
mechanical unfolding of 1TIT and a few about 1I6C have already been
reported in \cite{IPZ} and will be recalled here for comparison with
the other molecules. Results on the thermal unfolding of the other
molecules, based on the WS model, have been reported in \cite{Ap1}
(1COA) and \cite{BPZ1} (1BBL and 1I6C). In particular, in \cite{BPZ1} it
was shown that 1BBL differs from a clear two--state behaviour, though
not being a true downhill folder as some authors have claimed. In the
following we shall see that signals of the peculiar behaviour of 1BBL
appear also in the mechanical unfolding.

While the equilibrium properties of the present model can be calculated exactly  (see next section), results
on the unfolding kinetics will be instead obtained by performing Monte Carlo (MC) simulations with Metropolis algorithm, using the effective Hamiltonian
(\ref{hnoi}). In the following $t_0$ will indicate the system time
scale, corresponding to a single MC step.

\section{Equilibrium properties}\label{eq_p}
Once the temperature and the external force have been fixed, the
macroscopic state of the system is defined by the order parameter $m$
and by the molecule length $L$.

\begin{figure}[h]
\center
\psfrag{a}[ct][ct][1.]{(a)}
\psfrag{b}[ct][ct][1.]{(b)}
\psfrag{l2}[ct][ct][1.]{$\sqrt{\average{L^2}}\,  (\AA)$}
\psfrag{m}[ct][ct][1.]{$m$}
\psfrag{f}[ct][ct][1.]{$f$ (pN)}
\includegraphics[width=8cm]{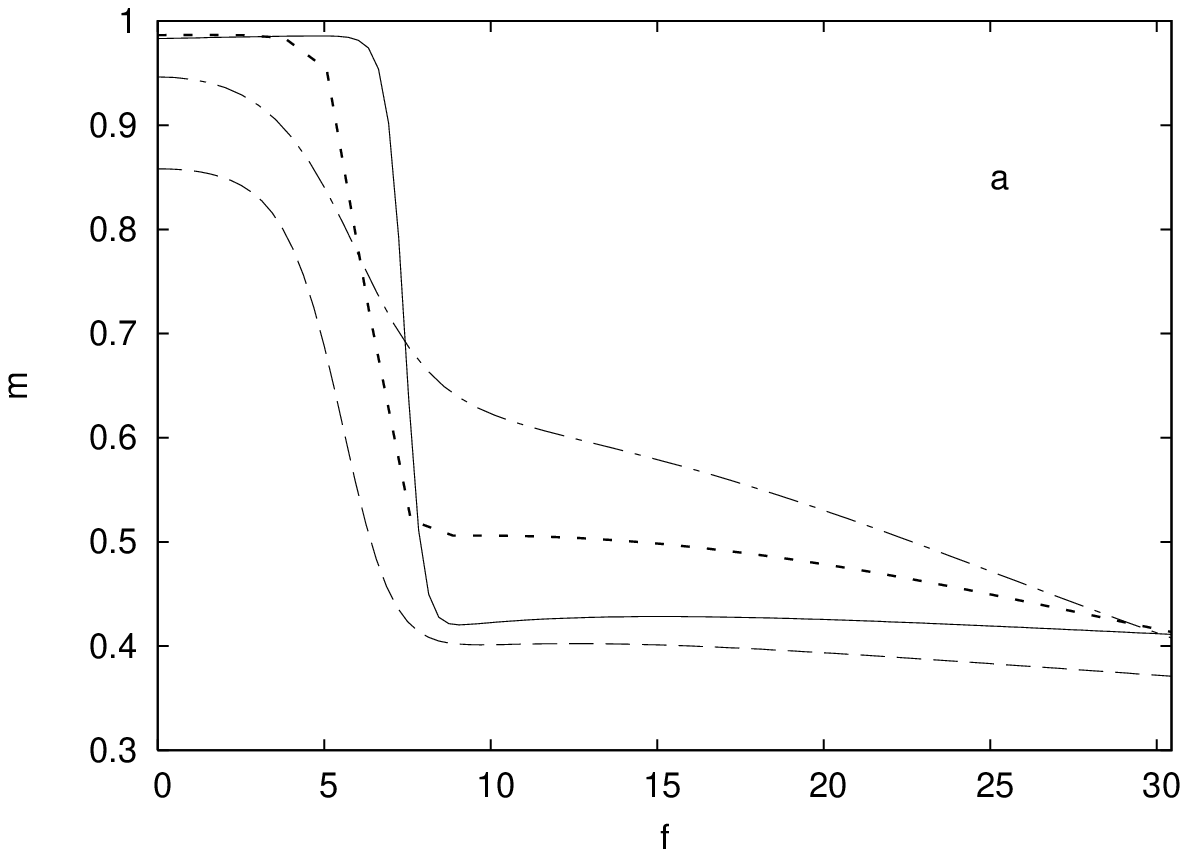}
\includegraphics[width=8cm]{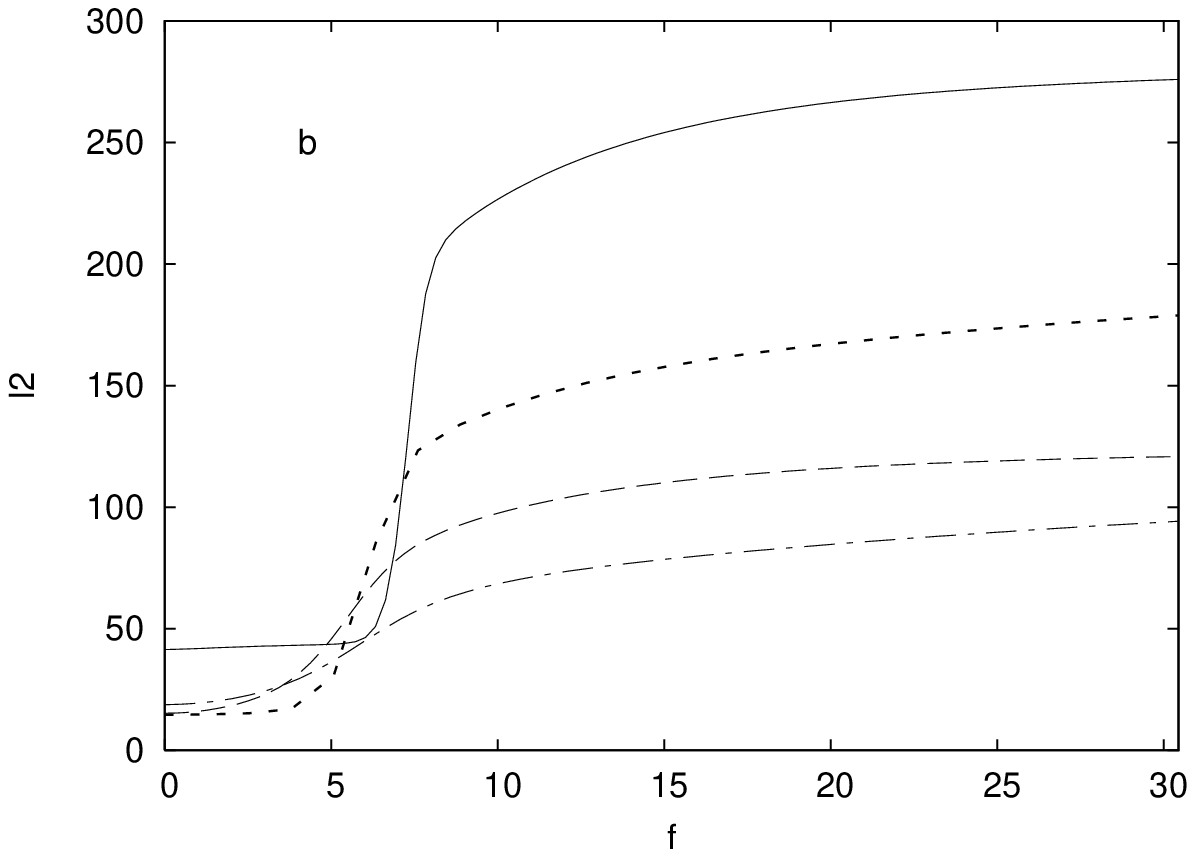}
\caption{Panel (a): average order parameter $m$ as a function of the
external force $f$, for different molecules: 1TIT (full line), 1COA
(dotted line), 1I6C (dashed line), 1BBL (dashed-dotted line). The
temperature value is taken to be $T=300$ K. Panel (b): Root mean
square length of the same molecules as a function of the external
force $f$, with $T=300$ K.}
\label{compare}
\end{figure}

In figure \ref{compare} the equilibrium values of such quantities are
plotted as functions of the force, at room temperature, for the
different molecules here considered.  Inspection of these figures
suggests that the two larger molecules exhibit a sharp transition to
the unfolded state, while for the two smaller molecules the unfolding is
more gradual as the force is increased.
The small force plateaux correspond to a global reorientation of the
molecule, which remains in its native state, under the external
force. When the force increases a more or less sharp transition to an
elongated state occurs. As in the thermal unfolding study \cite{BPZ1}
we see that 1BBL exhibits a different, more gradual transition with respect
to the other molecules, which exhibit a clearer two--state behaviour. 

\begin{figure}[h]
\center
\psfrag{f}[ct][ct][1.]{$f$ (pN)}
\psfrag{T}[ct][ct][1.]{$T$ (K)}
\includegraphics[width=9cm,height=7cm]{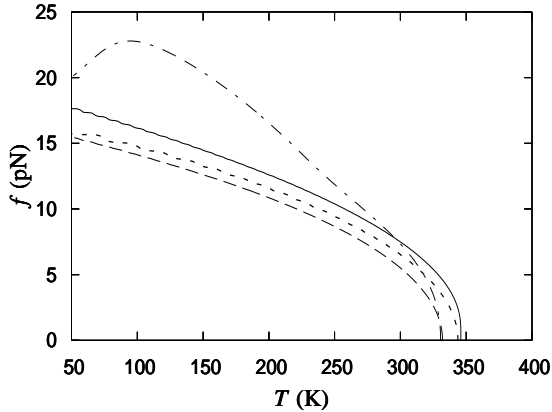}
\caption{Phase diagram of 
 1TIT (full line), 1COA (dotted line), 1I6C (dashed line), 1BBL
(dashed-dotted line). For each molecule, the curves are defined by $p(f,T)=1/2$.
The lower--left  region of the phase diagram corresponds to folded molecules, the upper-right region corresponds to unfolded molecules.}
\label{fase}
\end{figure}

For each molecule, we can obtain a phase diagram by computing the locus of points in the temperature--force plane where the
fraction of folded molecules $p=1/2$. Such curves are plotted in Fig.~\ref{fase}: inspection of this figure indicates, again,  that 1BBL differs qualitatively from the other molecules. 

\begin{figure}[h]
\center
\psfrag{F}[ct][ct][1.]{$F\, (k_B T)$}
\psfrag{L}[ct][ct][1.]{$L$ (\AA)}
\psfrag{a}[ct][ct][1.]{(a)}
\psfrag{b}[ct][ct][1.]{(b)}
\includegraphics[width=8cm]{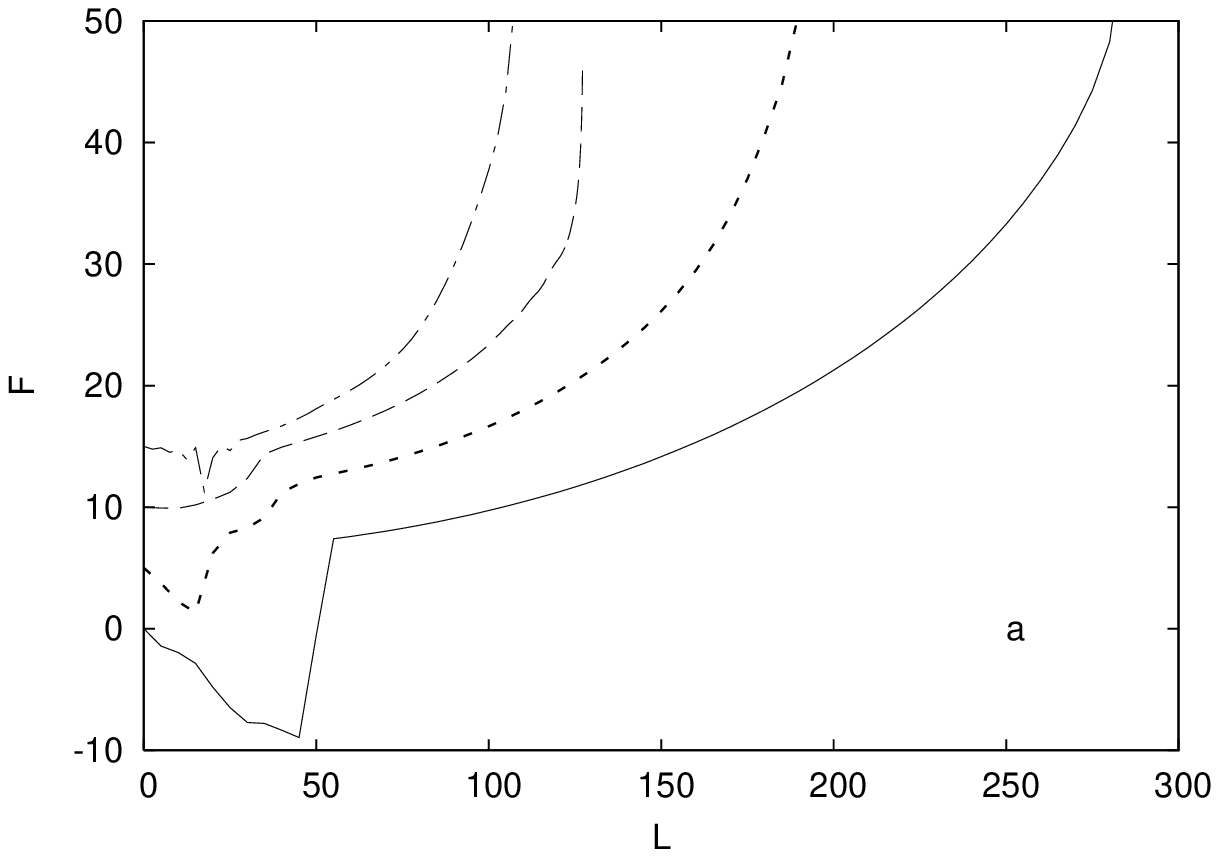}
\includegraphics[width=8cm]{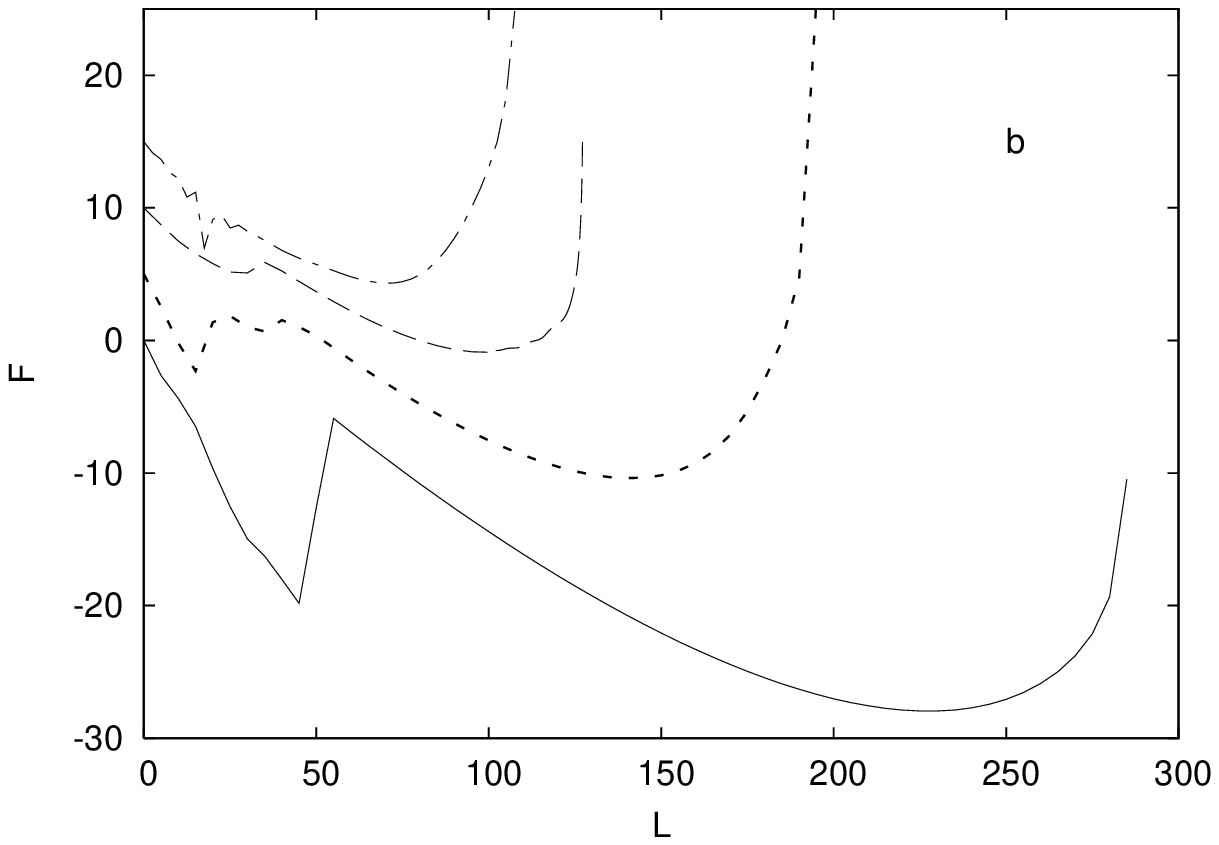}
\caption{Panel (a): Free energy landscape as a function of the
molecular elongation $L$, at $T=300$ K, for different molecules: 1TIT
(full line), 1COA (dotted line), 1I6C (dashed line), 1BBL
(dashed-dotted line). Panel (b): Tilted free energy landscapes
$F(L)-f\cdot L$, with $f=10$ pN.}
\label{comp_land}
\end{figure}

In order to characterize the macroscopic state of the molecules with
the observable quantity $L$, let us define the constrained zero--force partition
function $Z_0(L)$ as follows:
\begin{equation}
Z_0(L)=\sum_{\{m_k\},\{\sigma_{ij}\}} \delta(L-L(\{m_k\},\{\sigma_{ij}\}))\E^{-\beta \mathcal H(\{m_k\},\{\sigma_{ij}\},f=0)},
\label{zeta0}
\end{equation} 
and the corresponding free energy  by
\begin{equation}
F_0(L)=-k_B T \ln Z_0(L).
\label{effe0}
\end{equation} 
In appendix \ref{FEL} we show how $Z_0(L)$, as defined by eq.~(\ref{zeta0}), can be calculated exactly for our model.
In figure \ref{comp_land}(a), the equilibrium free energy landscape of
the four molecules here considered is plotted as a function of the molecule length,
for $T=300$ K.  Inspection of this figure indicates that all the
molecules have a minimum of the free energy at small $L$, the value of
the minimum being compatible with the value of $\sqrt{\average{L^2}}$
at zero force, as plotted in fig.~\ref{compare}(b).  At larger force, the
free energy function (\ref{effe0}) is tilted and reads $F(L,f) = F_0(L)
- f \cdot L$: this function exhibits new minima for each molecule,
which are compatible with the values of $\sqrt{\average{L^2}}$ in the
large force regime, see fig.~\ref{comp_land}(b).  As an example, let
us consider the 1TIT molecule. In fig.~\ref{comp_land}(b), the
function $F(L,f)$, with $f=10$ pN, exhibits a global minimum at
$L\simeq 225$ \AA, which corresponds to the equilibrium value of
$\sqrt{\average{L^2}}$ for the same molecule at  $f=10$ pN, as shown in
fig.~\ref{compare}(b). 

The function  $F(L,f)$ exhibits an energy barrier at small $L$,
whose 
 height $\Delta F$ and width $\Delta L$ depend on the value of $f$.
In order to give an example of the typical values of these quantities,
let us define $f_{1/2}$ as the force where the molecular length is half of its maximum value.

In table \ref{table_FL}, we list
the width $\Delta L$ and the height $\Delta F$ of the energy barrier
separating the two minima of the function $F(L)-f\cdot L$, for
$f=f_{1/2}$ and $T=300$ K.

\begin{table}[h]
\begin{tabular} {|c|c|c|}
\hline
Molecule & $\Delta L$ (\AA) & $\Delta F (k_B T)$ \\
\hline
1BBL   & 5 & 2.7\\
1I6C   & 10 & 1.6\\
1COA   & 25  & 5.7\\
1TIT   & 10  & 12.7\\
\hline
\end{tabular}
\caption{Width $\Delta L$ and height $\Delta F$ of the energy barrier
  separating the two minima of the free energy $F(L) = F_0(L) - f
  \cdot L$, with $f=f_{1/2}$, $T=300$ K, for the molecules here
  considered.}
\label{table_FL}
\end{table}

\section{Force Clamp}\label{FC}
In this section, we investigate the unfolding of the model proteins by
application of a constant external force: such a manipulation scheme
is usually called ``force clamp''.  The force on the molecule is
suddenly increased from 0 to its final value $f$, and its length is
measured \cite{exp_fc1, exp_fc2}.  Usually the unfolding of a small molecule or
of a portion of a large molecule is viewed as the overcoming of a
kinetic barrier in the molecular energy landscape
\cite{ev2}. Such a barrier is characterized by a width $x_u$ along the
reaction coordinate, and by a height $\Delta E_u$ over the
corresponding minimum.  Thus, the mean unfolding time is expected to
follow the Arrhenius law
\begin{equation}
\tau_u=\omega_0^{-1} \E^{\beta(\Delta E_u- f x_u)} = \tau_0
\E^{-\beta f x_u}, 
\label{tauf}
\end{equation} 
where $\omega_0$ is a microscopic attempt rate and $\tau_0 =
\omega_0^{-1} \exp(\beta \Delta E_u)$ is the mean unfolding time at
zero force.  Note that Eq.~(\ref{tauf}) is well defined as long as
$f\le \Delta \Eu/\xu$, i.e.  as long as the process can be actually
considered as a jump process over an energy barrier. For $f>\Delta
\Eu/\xu$ one expects that the mean escape time $\tu$ is independent of
the external force but rather depends on the microscopic details of
the system. On the other hand for too small forces the system will not
unfold, if the barrier $\Delta \Eu$ is sufficiently high ($\beta \Delta \Eu \gg 1$).

\begin{figure}[h]
\center
\psfrag{tau}[ct][ct][1.]{$\tau_u\, (t_0)$}
\psfrag{f}[ct][ct][1.]{$f$ (pN)}
\includegraphics[width=8cm]{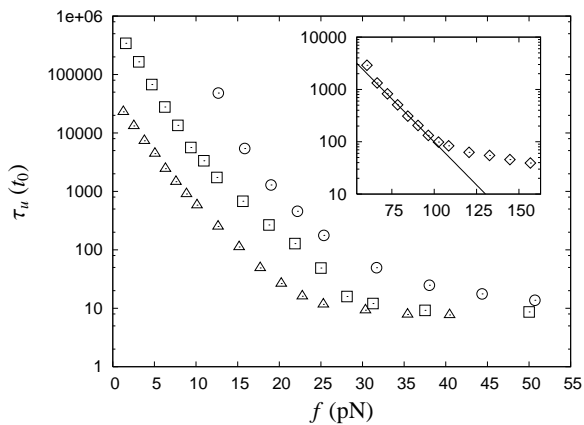}
\caption{Mean unfolding time as a function of the force, at $T=300$ K, for four different molecules: boxes 1I6C, triangles 1BBL, circles 1COA, inset: 1TIT.}
\label{comp_fc}
\end{figure}

In order to check whether the unfolding time under constant force
obeys such a law, and to extract the kinetic parameter $x_u$, we
run MC simulations to mimic the unfolding of molecules subject to a
force clamp with force $f$, at time $t=0$. We consider a molecule as
unfolded as soon as its length takes the value $L_u=L_{max}/2$, where
$L_{max}$ is the molecule length in the completely unfolded state. For
each molecule 1000 independent unfolding trajectories are considered.
In fig. \ref{comp_fc}, the mean unfolding time of the four molecules
is plotted as a function of the force. In the case of 1TIT, the force
 does not extend to the small force range, since we find that the unfolding time
$\tau_u$ goes to infinity in the very small force regime, i.e., the molecules
do not unfold at all. 
Inspection of this figure suggests that the mean unfolding time as a
function of the force follows eq. (\ref{tauf}) in a wide range of
force values, and then saturates in the large force regime, as found in other works \cite{ciep,LKH}.  The force value separating the two regimes depends on the
particular molecule.  From fits to eq.~(\ref{tauf}) the values
of the kinetic parameter $x_u$ can be obtained for each
molecule. In table \ref{table_fc} the values of the unfolding length are listed for the molecules considered
in this paper. In principle, from such a fit procedure one is also able to obtain an estimate of $\tau_0$. However the quantity $\tau_0$ cannot be expressed in seconds, since this would require to evaluate the molecular time scale $t_0$. This, in turn,  requires an  experimental estimate of $\tau_0$ with the force clamp technique.
It is worth noting that, despite the simplicity of our model,  the
unfolding length for 1TIT is  in remarkable agreement with the experimental value $\xu=2.5\, \AA$~\cite{rgo}.
\begin{table}[h]
\begin{tabular} {|c|c|}
\hline
Molecule & $x_u$ (\AA)   \\
\hline
1BBL   & $14.8 \pm 0.4$ \\
1I6C   & $20.3 \pm 0.5$ \\
1COA   & $20\pm 2 $  \\
1TIT   & $3.13\pm 0.1$  \\
\hline
\end{tabular}
\caption{Unfolding length $x_u$  as given from linear fits to eq.~(\ref{tauf}), for the molecules considered in this paper.}
\label{table_fc}
\end{table}

\begin{figure}[h]
\center
\psfrag{a}[ct][ct][1.]{$(a)$}
\psfrag{b}[ct][ct][1.]{$(b)$}
\psfrag{tu}[ct][ct][1.]{$\tu\, (t_0)$}
\includegraphics[width=8cm]{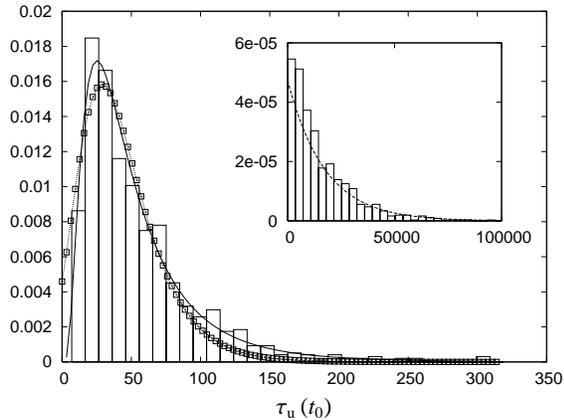}
\caption{Histograms of the unfolding time $\tu$ for the 1TIT molecule
in a force clamp, at $T=300$ K with force $f=130$ pN (main figure) and
$f=48$ pN (inset).  The lines in the main figure are fits of the data to a log-normal
function (full line) and  to the function (\ref{ftau}) (line-points). The line in the inset corresponds to a fit  to a negative exponential function.}
\label{fc_histo}
\end{figure}

The unfolding of the molecule is a stochastic process, and thus the
unfolding time varies for each realization of the process. It is thus
interesting to study the distribution of the unfolding time.  In
figure \ref{fc_histo}, we plot histograms of the unfolding time of the
1TIT molecule at $T=300$ K, for small and large force. In the large
force regime a lognormal distribution was proposed in \cite{ciep},
without a theoretical justification, and shown to fit quite well the
results of molecular dynamics simulations. Our results are also well
fitted by a lognormal distribution. Nevertheless, we find it
interesting to look for a theoretical distribution for the large force
regime. 

In the large force regime, one can imagine the chain as made up of a
number $M<N$ of stretches which can be easily turned by the force from
the antiparallel to the parallel (with respect to the force itself)
configuration. At the beginning of the pulling process, these
stretches will be randomly oriented, half of them antiparallel and 
 half parallel to the force. With a frequency $\tau_1^{-1}$, a
stretch will be selected at random by the kinetics and, if
antiparallel to the force, turned parallel with probability
1. Therefore, after a time $\tau_u = k \tau_1$, the probability that
the chain will be completely elongated in the force direction will be
$p(M,k) = [1 - (1 - 1/M)^k]^{M/2}$. The probability distribution of
the unfolding time will therefore be approximated by $f(\tau_u = k
\tau_1) = p(M,k) - p(M,k-1)$, which for large $M$ can be approximated
as 
\begin{equation}
f(\tau_u=k \tau_1)=1/2 \exp[ -k/M - M/2 \exp(-k/M)]. 
\label{ftau}
\end{equation} 
As can be seen in Fig.~\ref{fc_histo}, this formula  fits reasonably well our data, though
not as well as a lognormal. Similar results are obtained for the other
molecules (data not shown).

\subsection{The Kramers problem}
We now ask the following question: is the free energy landscape $F(L)$
the ``kinetic'' potential associated to the diffusive motion of the
system along the coordinate $L$? In other words, in the zero force
regime, is the thermal motion of the molecule a diffusion process
across the potential $F_0(L)$? A positive answer to this question would
imply that the mean first passage time of the system to a given value
$L^*$ of the elongation should be given by the solution of the Kramers
problem.  The Kramers problem amounts to evaluating the mean first
passage time (mfpt) of a Brownian particle, moving in an energy
potential $U(x)$, to a given point $x_f$ of the potential.  If we let $x_0$
be the initial position of the particle, the mfpt from $x_0$ to $x_f$
reads \cite{Zwa}
\begin{equation}
\tau(x_f)=\frac 1 D\int_{x_0}^{x_f} d y \E^{\beta U(y)} \int_{-\infty}^y \E^{-\beta U(z)},
\label{tau_kra}
\end{equation} 
where  $D$ is the diffusion coefficient, which basically
sets the time scale of the process.

\begin{figure}[h]
\center
\psfrag{a}[ct][ct][1.]{(a)}
\psfrag{b}[ct][ct][1.]{(b)}
\psfrag{f}[ct][ct][1.]{$f$ (pN)}
\psfrag{t}[ct][ct][1.]{mfpt ($t_0$)}
\psfrag{F}[ct][ct][1.]{$F\, (k_B T)$}
\psfrag{L}[ct][ct][1.]{$L$ (\AA)}
\psfrag{f0}[ct][ct][.8]{$f=0$}
\psfrag{f5}[ct][ct][.8]{$f=5$}
\psfrag{f15}[ct][ct][.8]{$f=15$}
\psfrag{T6.6}[ct][ct][.8]{$T=300$}
\psfrag{T6}[ct][ct][.8]{$T=272$}
\psfrag{T4}[ct][ct][.8]{$T=181$}
\psfrag{L2}[cl][cl][.75]{$L_{max}/2$}
\includegraphics[width=8cm]{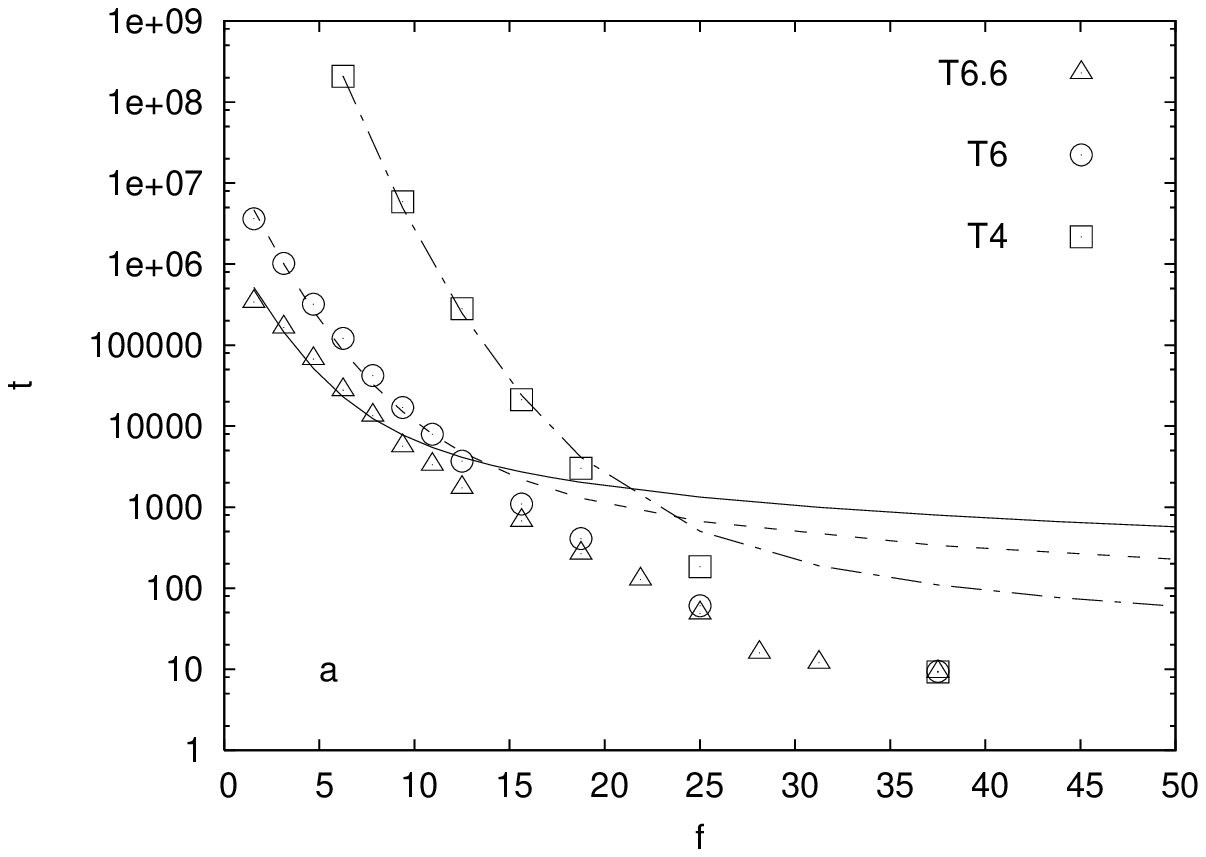}
\includegraphics[width=8cm]{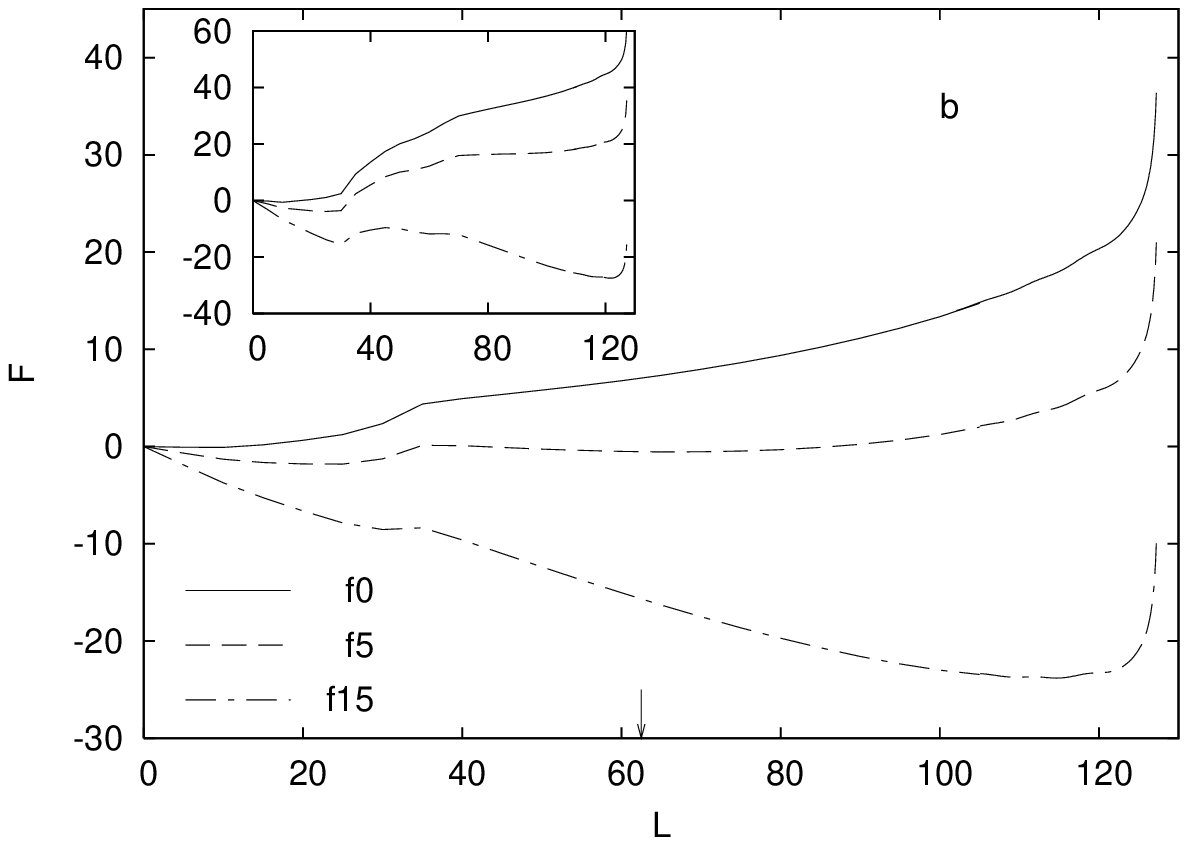}
\caption{Panel (a): Mean first passage time at $L^*=L_{max}/2$ for the
1I6C molecule and for different temperatures (in K), as obtained by direct computer simulations (points) and by
eq.~(\ref{tau_kra}) (lines). Panel (b): Free energy landscape $F$ as a function of the
molecule length $L$, at $T=300$ K, for different values of the force (in pN).
Inset: Free energy landscape $F$ at $T=181$ K.}
\label{tau_1I6C}
\end{figure}

In figure~\ref{tau_1I6C}(a), we plot the mean first passage time of
the 1I6C molecule at $L^*=L_{max}/2$, as a function of $f$, for different
temperatures, as obtained by eq.~(\ref{tau_kra}), where $U(x)$ has
been replaced by the free energy $F_0(L)-f\cdot L$, as given by
eq.~(\ref{effe0}).  Inspection of this figure suggests that in the
small force range, the unfolding kinetics of the protein is actually
described by the equilibrium free energy landscape $F(L,f)$. The
agreement improves as the temperature decreases. The mean first
passage time of the molecule is not recovered by eq. (\ref{tau_kra})
in the large force regime, because the energy difference
$F(L^*,f)-F(L=0,f)$ becomes negative, and thus the motion towards the new
minimum at large $L$ becomes purely diffusive, see
fig.~\ref{tau_1I6C}(b).

The same behaviour is found for the 1BBL molecule (data not shown). Since the unfolding
time grows exponentially with the size of the molecule, for the two
larger molecules we were not able to observe the unfolding in the
small force regime within reasonable computation times.

\section{Dynamic loading}\label{DL}

In this section we consider the following manipulation strategy, which
is often used in experiments \cite{KSGB,rgo,cv1,DBBR,Ober1}: starting from equilibrium
configurations, a time-dependent force is applied to our model
molecule and the unfolding time is sampled.  As in the case of the
force clamp, we define the unfolding time as the first passage time of
the molecule length across the threshold value $\Lu$.  Here the force
is taken to increase linearly with time, with a rate $r$: this manipulation scheme corresponds to the force--ramp experimental set--up discussed in \cite{Ober1}.  Thus, the
rupture force $f_u$ is given by $f_u=r \tu$.

\begin{figure}[h]
\center
\psfrag{a}[ct][ct][1.]{(a)}
\psfrag{b}[ct][ct][1.]{(b)}
\psfrag{f}[ct][ct][1.]{$f^*$ (pN)}
\psfrag{v}[ct][ct][1.]{$r\, (\mathrm{pN}/t_0)$}
\psfrag{T6}[cc][cc][.8]{$T=262$}
\psfrag{T7}[cc][cc][.8]{$T=300$}
\psfrag{T4}[cc][cc][.8]{$T=175$}
\includegraphics[width=8cm]{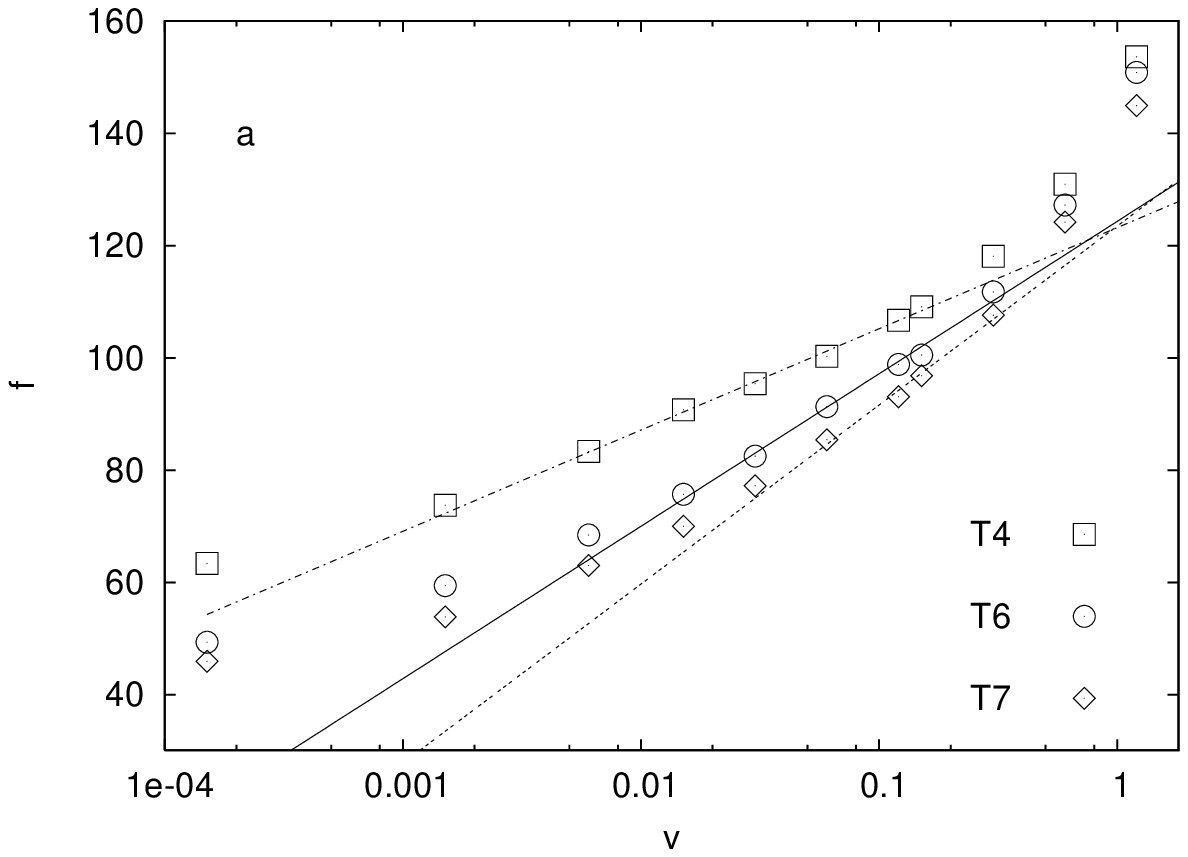}
\includegraphics[width=8cm]{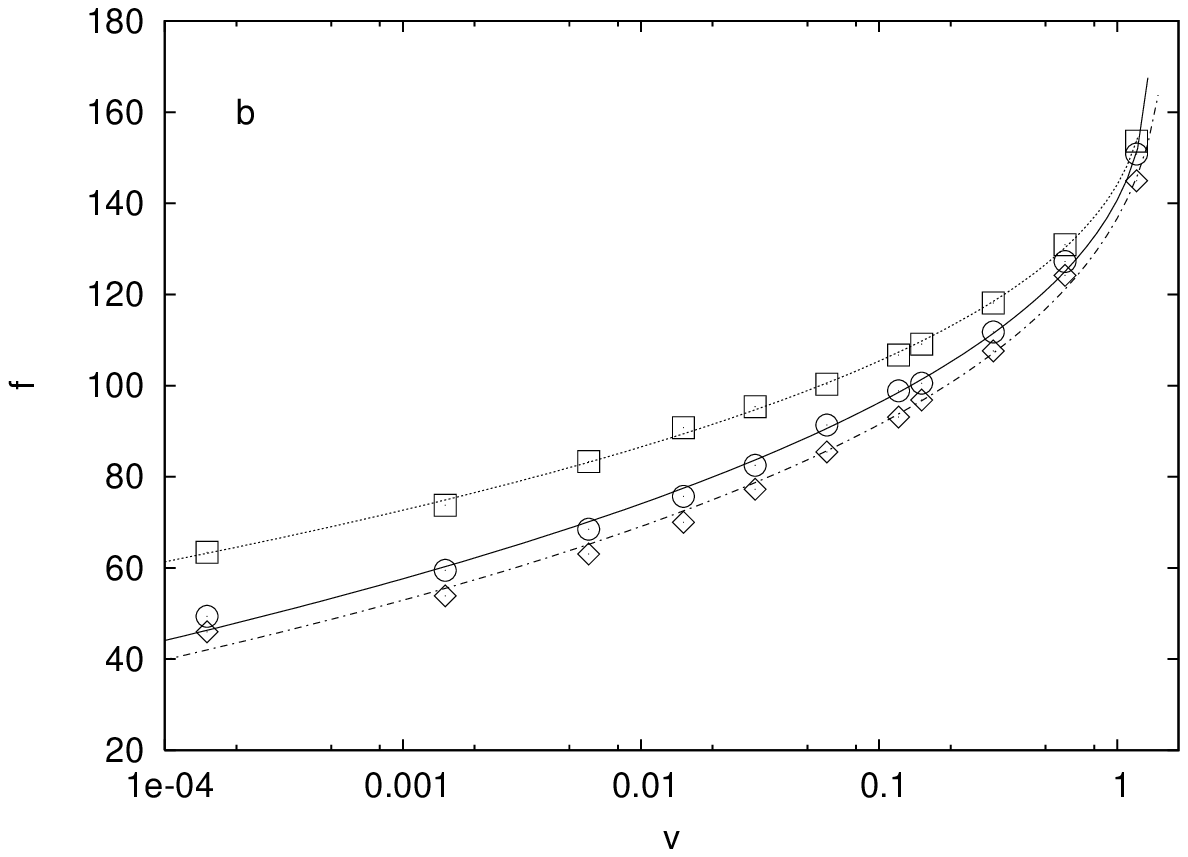}
\caption{Panel (a): plot of the typical unbinding force $f^*$ of the
1TIT molecule as a function of the pulling velocity $r$, for the three
values of the temperature. The lines are fits to the data in the
linear regime defined by eq.~(\ref{fstar}). From such fits one can
obtain the characteristic unfolding length $x_u$.  We find
$x_u=3.1\pm0.1\, \AA$ for $T=175$ K, $x_u=3.0\pm0.1\, \AA$ for
$T=262$ K, and $x_u=3.05\pm 0.2\, \AA$ for $T=300$ K. Panel (b): the
lines are fits of the data to eq.~(\ref{hs_eq}). }
\label{dl_1TIT}
\end{figure}

As discussed above, the breaking of a molecular linkage is typically
described as a thermally activated escape process from a bound state
over a barrier which dominates the kinetics. It can be shown that, if
the energy barrier $\Delta E_u$ is large (compared to the thermal
energy $k_B T$) and rebinding is negligible, the typical unbinding
force of a single molecular bond reads \cite{ev2,denis2}
\begin{equation}
f^*=\frac{k_B T}{x_u} \ln \pq{\frac{r  x_u\tau_0}{ k_B T}}.
\label{fstar}
\end{equation} 
 In
fig.~\ref{dl_1TIT}(a) the typical unbinding force $f^*$ (the most
probable value of $f_u$) is plotted as a function of the pulling
velocity for the 1TIT molecule, for three values of the temperature.
Inspection of this figure suggests that the range of values of $r$,
where $f^*$ is a linear function of $\ln r$, depends on the
temperature: the smaller is $T$, the wider is this range.  We find
that, for this molecule, the value of the unfolding length, $x_u\simeq
3\, \AA$, obtained by fitting the data in the linear regime to eq.~(\ref{fstar}), is
independent of the temperature, as expected (see caption of the figure
for the numerical values). Note that the value of the unfolding
length, $x_u$, in the linear regime defined by eq.~(\ref{fstar}),
agrees with that found with the force clamp manipulation, and with the
experimental value $\xu=2.5\, \AA$ found in ref.~\cite{rgo}.

\begin{table}[h]
\begin{tabular} {|c|c|}
\hline
Molecule & $x_u$ (\AA) \\
1BBL   & $14.0 \pm 0.3$ \\
1I6C   & $17.7 \pm 0.5$ \\
1COA   & $22.4\pm 0.5$  \\
1TIT   & $3.05 \pm 0.20 $ \\
\hline
\end{tabular}
\caption{Unfolding length $x_u$,  as given from linear fits to eq.~(\ref{fstar}), for the
molecules considered in this paper.}
\label{table_dl}
\end{table}

By repeating the same procedure for the other molecules, we estimate
their unfolding lengths, the results are listed in table
\ref{table_dl}. Again, $\tau_0$
cannot be directly compared with experimental values, but we see that
the unfolding lengths are comparable to the ones obtained by the force
clamp protocol.


However, as discussed above the rupture force is not a linear function 
of $r$ in the whole range of the pulling rate: figure \ref{dl_1TIT}  rather suggests that the slope of $f^*$ increases as $r$ increases.
In refs.~\cite{denis1,denis2,alb2} it has been argued that the appearance of
different slopes in the $f^*$ vs. $\ln r$ plot could be the signature of the
presence of different escape paths from the folded state.  Each of
these different paths would be selected by pulling the molecule with a given rate $r$.
 Thus, the different slopes in the $f^*$ vs. $\ln r$ plot correspond
to different characteristic lengths $x_u$ of the paths.  On the other
hand, in a recent work \cite{hs_paper}, considering particular
choices of the energy landscape which make exact computations
feasible, it has been argued that, the typical unbinding force $f^*$
has a more complex expression
\begin{equation}
f^*=\frac{\Delta E_u}{\nu x_u}\pg{1-\pq{\frac {k_B T}{\Delta E_u} \ln\p{\frac{\omega_0 \E^\gamma}{\beta x_u r}}}^\nu},
\label{hs_eq}
\end{equation} 
where the exponent $\nu$ depends on the microscopic details of the
energy landscape, and $\gamma$ is the Euler-Mascheroni constant
$\gamma\simeq0.577$.  Equation~(\ref{hs_eq}) reduces to
eq.~(\ref{fstar}) in the limit $\Delta E_u \rightarrow \infty$, or
when the exponent $\nu$ takes the value 1 \cite{hs_paper}.
A similar expression for the rupture force was previously proposed in \cite{Dudk}, with $\nu=2/3$.
In fig.~\ref{dl_1TIT}(b), the fits of the typical unbinding force data
to eq.~(\ref{hs_eq}) are plotted for the 1TIT molecule. The fits turn out to be
rather good, but statistical errors are quite large. In order to
reduce them, for each molecule, we considered sets of $(r,f^*)$ data
at different temperatures, and we made joint fits according to
Eq.~(\ref{hs_eq}).  The values of the unfolding lengths, energy
barriers and exponents obtained by these fits are listed in table
\ref{table_fstar}.
\begin{table}[h]
\begin{tabular} {|c|c|c|c|}
\hline
Molecule & $x_u$ (\AA)  & $\Delta E_u\, (k_BT,\, T=300{\mathrm K})$ & $\nu$\\
\hline
1BBL   & $22\pm 1$ & $10.0\pm 0.2$ &$0.61\pm 0.03$\\ 
1I6C   & $18\pm1$ & $13\pm 2$ & $0.7\pm 1$\\
1COA   & $31.5\pm 4 $ &$16\pm 1$ & $0.625\pm 0.06$  \\ 
1TIT   & $11.5\pm 2$ &$18\pm 1$ & $0.42\pm0.04$\\
\hline
\end{tabular}
\caption{Unfolding length $x_u$, barrier height $\Delta E_u$, and characteristic exponent $\nu$ as given from fits of the unfolding data with dynamic loading technique to
eq.~(\ref{hs_eq}).}
\label{table_fstar}
\end{table}
Comparison of the unfolding lengths listed in table \ref{table_fstar}
and in table \ref{table_dl} indicates that the values of the $x_u$
obtained by fitting the typical unbinding force to eq.~(\ref{hs_eq})
are rather different from the values obtained by fitting the same data
to eq.~(\ref{fstar}), although, as discussed before, the fit to
eq.~(\ref{fstar}) is restricted to the intermediate range of
values of $r$. This is  in agreement with
refs. \cite{Dudk,hs_paper}. In those references it was argued that eq.~(\ref{hs_eq}) describes the rupture force in the whole regime of $r$
rather than just in the linear regime, as eq.~(\ref{fstar})
does. 
It is worth noting that the values of the exponent $\nu$ found for the 
1BBL, 1I6C and 1COA molecules (table 
\ref{table_fstar}) are compatible with the value $\nu=2/3$ found 
in refs.~\cite{Dudk, hs_paper} for a particular choice of the energy landscape.
Furthermore, the value of the kinetic barrier $\Delta E_u=18\pm 1\, k_B T$ (at $T=300$ K) found for the 1TIT molecule, is of the same order of magnitude of the experimental value $\sim 37.3 k_B T$ found in \cite{rgo}.

Similarly to the case of the force clamp, the distribution of the
unbinding force exhibits a nontrivial dependence on the system kinetic
parameters \cite{ev1}:
\begin{equation}
P(f)=\frac{1}{\tau_0 r} \E^{\beta f x_u} \exp\pq{-\frac{k_B T}{r x_u\tau_0} \p{\E^{\beta f x_u}-1}}.
\label{p_f}
\end{equation}
The maximum of this distribution corresponds to the typical unbinding
force, eq.~(\ref{fstar}).

\begin{figure}[h]
\center
\psfrag{f}[ct][ct][1.]{$f$ (pN)}
\psfrag{a}[ct][ct][1.]{(a)}
\psfrag{b}[ct][ct][1.]{(b)}
\includegraphics[width=8cm]{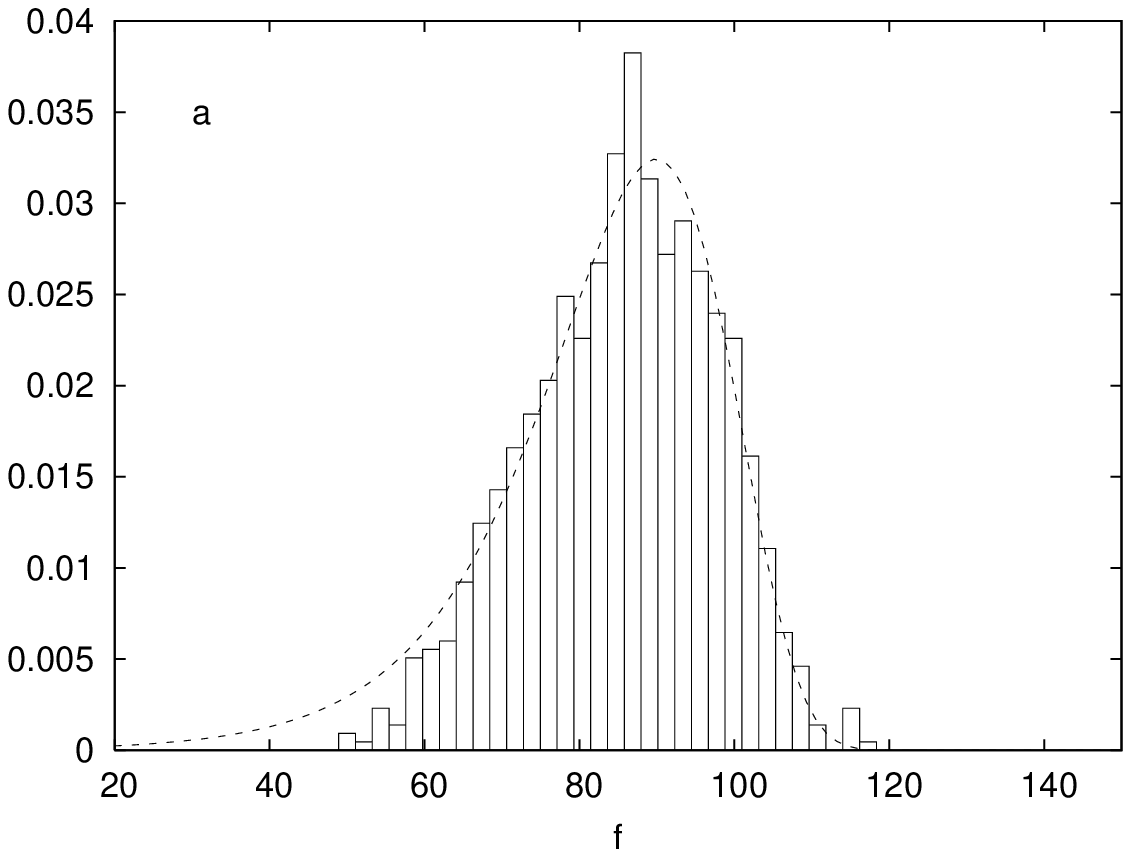}
\includegraphics[width=8cm]{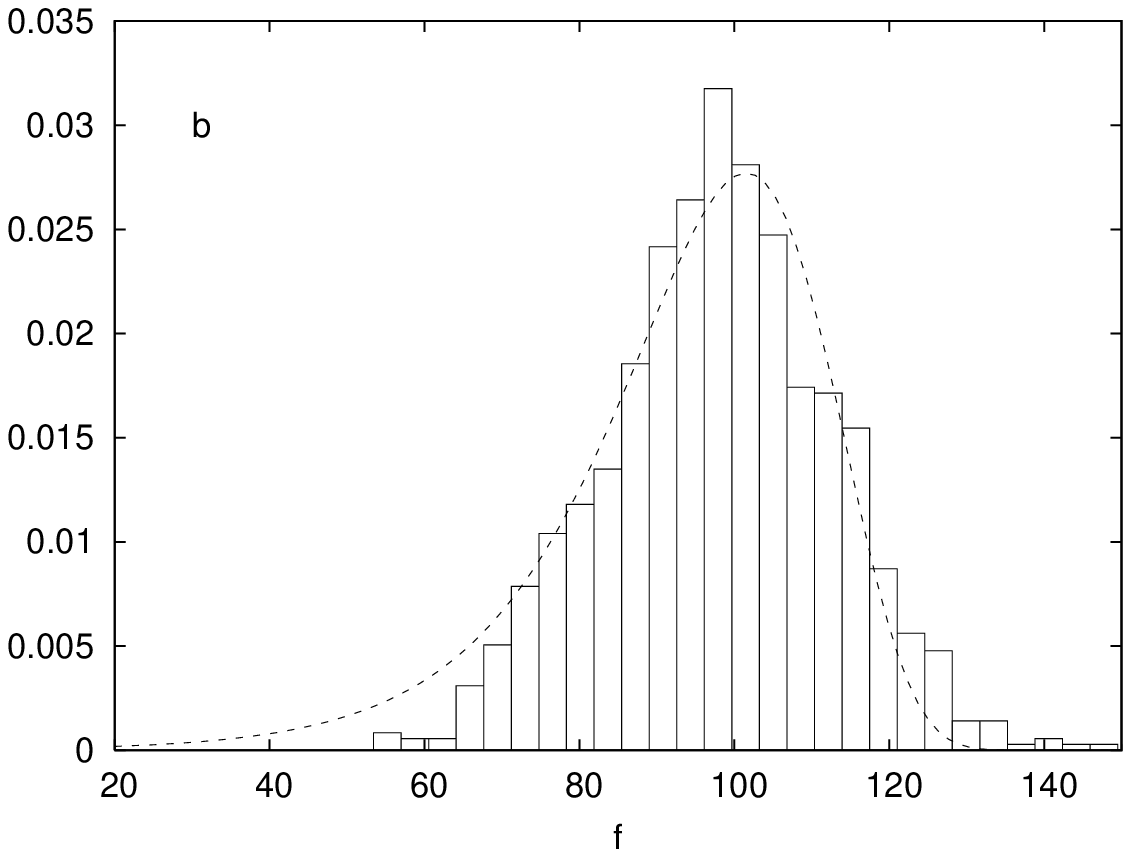}
\caption{Distribution of unfolding force under dynamical loading for
the 1TIT molecule for (a) $r=0.06$ pN/$t_0$  and (b) $r=0.15$ pN/$t_0$, with $T=300$ K. The dotted line is a fit to eq. (\ref{p_f}). }
\label{histo_dl}
\end{figure}

In figure \ref{histo_dl}, the distribution of the unfolding force is
plotted for the 1TIT molecule and different rate values.  Inspection
of fig.\ \ref{histo_dl} suggests that the observed distribution of
unfolding force agrees nicely with the expected one. Similar results
are obtained for the other molecules. 

\begin{figure}[h]
\center
\psfrag{a}[ct][ct][1.]{(a)}
\psfrag{b}[ct][ct][1.]{(b)}
\psfrag{xu}[bc][bc][1.]{$x_u$ (\AA)}
\psfrag{r}[ct][ct][1.]{$r\, (\mathrm{pN}/t_0)$}
\includegraphics[width=8cm]{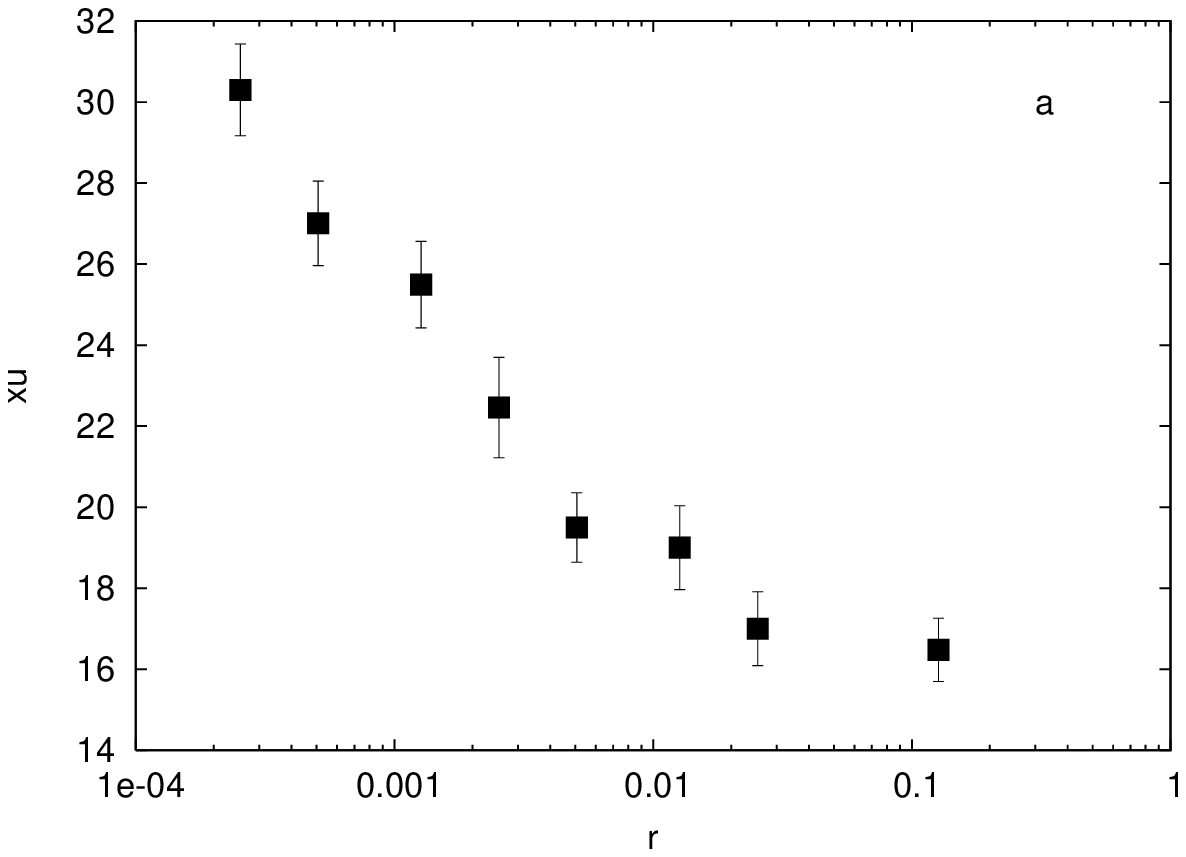}
\includegraphics[width=8cm]{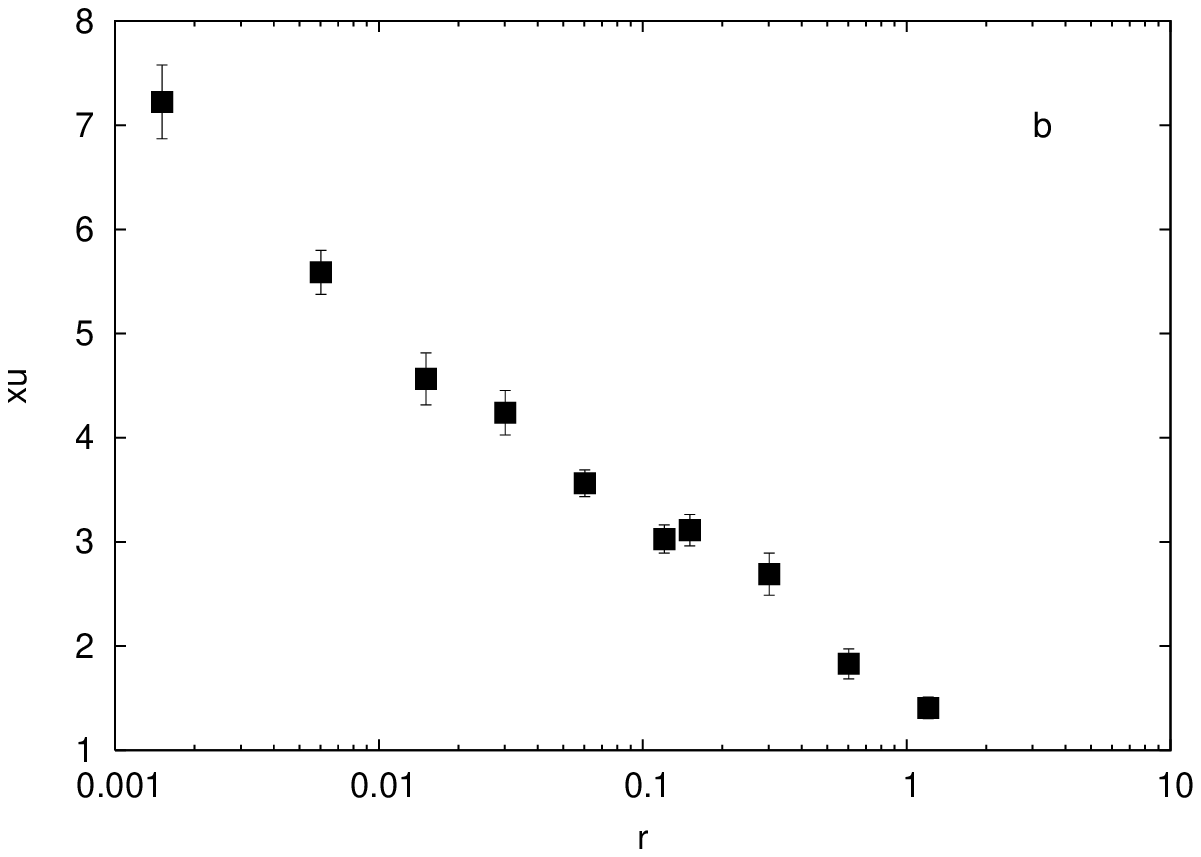}
\caption{Unfolding length $x_u$ as a function of $r$, as obtained from
fits of the unfolding force probability distribution to
eq.~(\ref{p_f}) with $T=300$ K, for the 1COA (a) and 1TIT (b) molecules.}
\label{xu_dl}
\end{figure}

In order to characterize the different unfolding paths occurring for
different values of $r$, one can assume that the unfolding length
$x_u$ is a function of the pulling rate $r$, and exploit
eq.~(\ref{p_f}) to extract the value $x_u(r)$, so as to estimate the
unfolding length at any value of $r$.  In figure \ref{xu_dl} the
unfolding length $x_u$ of the 1COA and 1TIT molecules are plotted as a function
of $r$, as obtained from this fitting procedure.  As expected we find
that $x_u$ is a decreasing function of $r$.  Thus, the values
$x_u=22.4$ \AA (1COA) and $x_u=3.05$ \AA (1TIT), obtained by fitting the typical unbinding force to
eq.~(\ref{fstar}) (see, table \ref{table_dl}), turn out to be 
weighted averages of the values $x_u(r)$ as plotted in
fig.~\ref{xu_dl}.

\section{Evaluating the free energy landscape from pulling
  experiments}\label{EJE} 

Given that we can compute exactly the free energy landscape $F_0(L)$ of
our model, let us now address the problem of evaluating this landscape
through the force manipulation experiments. As discussed above,
applying a dynamic loading is equivalent to drive the system out of
equilibrium by coupling it to the external potential $U(L,t)=-f(t) L$.

The free energy landscape can be evaluated
by using a fluctuation relation, which is an extended form of the Jarzynski equality \cite{HumSza,alb1}.
\begin{equation}
  \average{\delta(L-L(\{m_k\},\{\sij\}))\E^{-\beta W}}_t
  =\E^{-\beta \left(F_0(L)-f(t) L\right)}/Z_0,
\label{sample}
\end{equation}
where $Z_0=\int d L\,  Z_0(L)$, with $Z_0(L)$ as given by eq.~(\ref{zeta0}).
Combining the Jarzynski equality \cite{jarz}, and
the weighted histogram method \cite{ferr}, it can be shown that, if the
molecule is driven out of equilibrium by a time-dependent external
potential coupled to its length $U(L,t)$, the free energy $F$ as a
function of $L$ is given by \cite{HumSza,alb1}
\begin{equation}
F(L)=-k_B T \ln \pq{\frac{\sum_t\frac{\average{\delta(L-L(\{m_k\},\{\sij\})) \exp\p{-\beta W_t}}_t}{\average{ \exp\p{-\beta W_t}}_t}}{\sum_t\frac{\exp\p{-U(L,t)}}{ \average{\exp\p{-\beta W_t}}_t}}}, 
\label{ext_jarz}
\end{equation} 
where $W_t$ is the thermodynamic work {\it done} on the system by the external potential,
up to the time $t$, defined as $W_t=\int^t_0 d t'\partial \mathcal H/\partial t'  $, and the average $\average{\dots}_t$ is over all
the trajectories of fixed duration $t$.  In an experimental situation, the work $W_t$ is not sampled continuously, but at successive discrete times $0,\Delta t, 2 \Delta t,\dots, M\Delta t$.  Therefore the sum over $t$ in Eq.~(\ref{ext_jarz}) runs over these discrete values.

\begin{figure}[h]
\center
\psfrag{a}[ct][ct][1.]{(a)}
\psfrag{b}[ct][ct][1.]{(b)}
\psfrag{F}[ct][ct][1.]{$F\, (k_B T)$}
\psfrag{L}[ct][ct][1.]{$L\, (\AA)$}
\psfrag{1e2}[cl][cl][.8]{$r=0.6$}
\psfrag{1e3}[cl][cl][.8]{$r=0.06$}
\psfrag{1e4}[cl][cl][.8]{$r=6\cdot 10^{-3}$}
\psfrag{1e5}[cl][cl][.8]{$r=6\cdot 10^{-4}$}
\psfrag{2.5e-4}[cl][cl][.8]{$r=2.5\cdot 10^{-4}$}
\psfrag{2.5e-1}[cl][cl][.8]{$r=2.5\cdot 10^{-1}$}
\psfrag{5e-2}[cl][cl][.8]{$r=5\cdot 10^{-2}$}
\includegraphics[width=8cm]{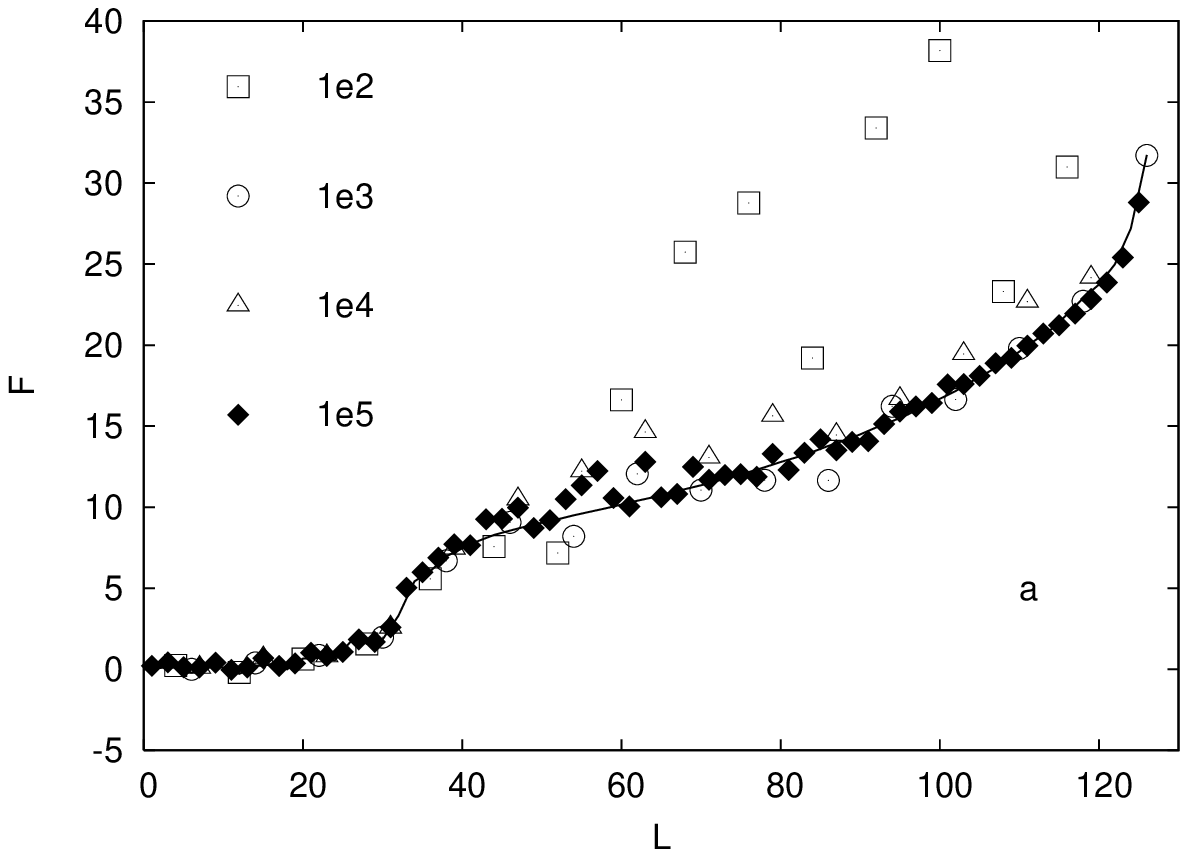}
\includegraphics[width=8cm]{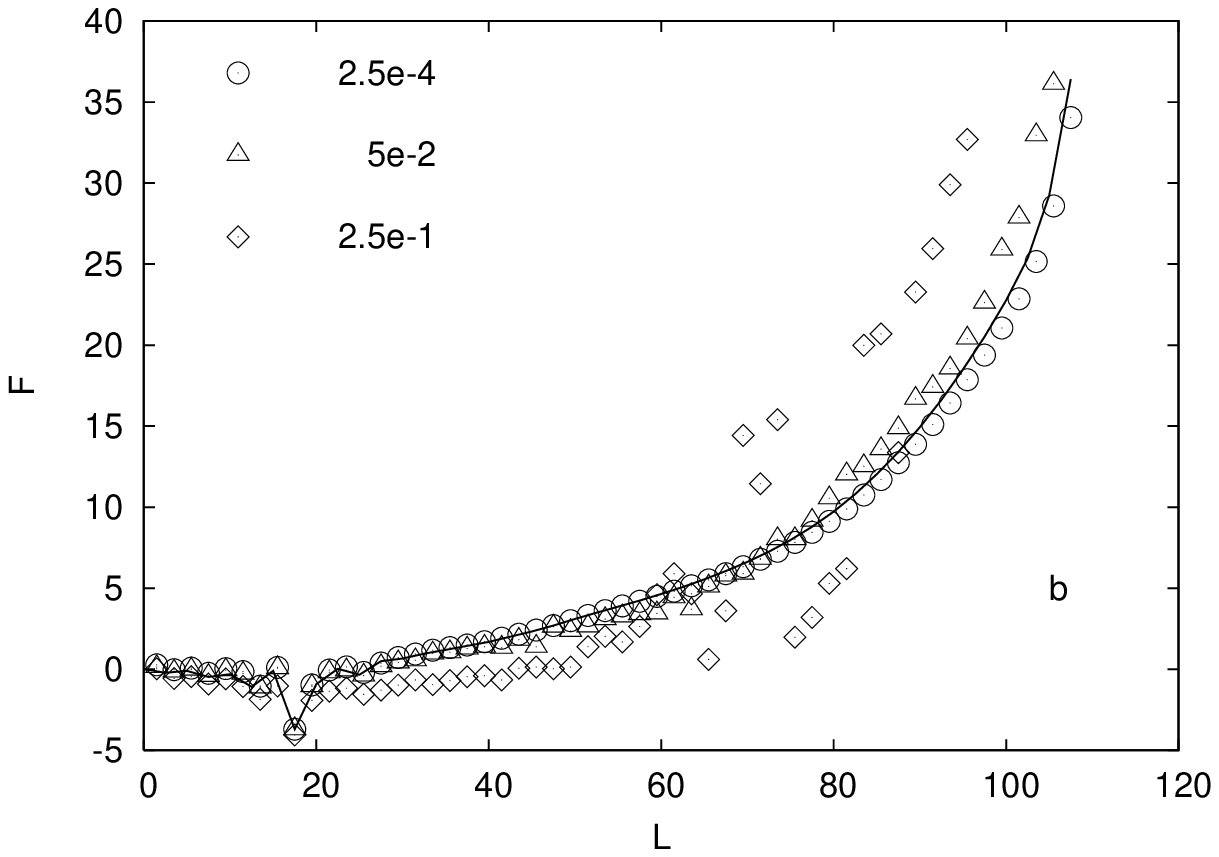}
\caption{Reconstructed free energy landscape $F$ of the PIN1 (a), and the 1BBL (b) for $T=300$ K.
The lines correspond to the exact result.
}
\label{land_jarz}
\end{figure}

The estimated free energy for the smaller molecules are plotted in
fig.~\ref{land_jarz}, as obtained by 10000 independent pulling
trajectories, together with the exact ones.  As expected
\cite{IPZ,alb1}, the curves obtained by numerical ``experiments"
collapse onto the expected one as the pulling rate $r$ decreases.

Within the present scheme, it was not possible to evaluate the free energy landscape of the two larger molecules:
indeed the work needed to completely unfold the two molecules (1COA, 1TIT) amounts to some hundreds of $k_B T$, and thus 
the numerical precision in evaluating the average value of $\exp(-\beta W)$ is rather scanty.

Our results suggest a practical procedure to estimate the free energy landscape
of real proteins with dynamic loading experiments.
The work done on the molecule has to be sampled for different pulling rates:
for each rate eq.~(\ref{ext_jarz}) provides an estimate of the target function $F_r(L)$.
As the rate $r$ decreases, the curves are expected to superimpose more
and more: when the difference between the curves is of the order of
few $k_B T$, for the whole range of $L$, the estimate of $F(L)$ can be
considered reliable within this small uncertainty.

\section{Conclusions}\label{Conclusions}
In this paper we presented a comparative study of four widely studied proteins, by exploiting a simple model recently introduced.
Such a model allows one to obtain analytically the equilibrium properties
of the molecules considered. 
By using MC simulations we also study the unfolding  kinetics of the molecules
as they are pulled through an external force applied to their free ends.
As already discussed in \cite{IPZ}, the model turns out to exhibit the typical
behaviour of a protein undergoing a mechanical manipulation, both with the 
force--clamp scheme and with the dynamic--loading scheme.

We systematically study the {\it equilibrium} free energy landscape of the model molecule, as a function of an experimentally accessible coordinate, namely the molecular elongation.
By comparing the unfolding time with the expected values, as given by the Kramers' formula, we find that, in the limit of small forces, the free energy landscape (\ref{effe0}) represents the {\it kinetic} energy landscape which governs the unfolding kinetics of the molecules, as discussed in ref.~\cite{ev1}.
Furthermore, by comparing the typical width of the energy barrier, as found from eq.~(\ref{effe0}) and listed in table \ref{table_FL}, with the unfolding length obtained from out-of-equilibrium unfolding experiments,  listed in tables \ref{table_fc}--\ref{table_dl},
we conclude that the latter parameters are not related to any typical length in the molecules, but are rather effective parameters.

By considering pulling at different velocities, we found that the unfolding length varies with the pulling velocity: thus our results support the point of view that different escape paths exist for a given molecule, each path being selected by the features of the manipulation.

Finally, the possibility of computing the free energy landscape as an exact result, make our model an excellent test bed to check the application of the fluctuation relations to the study of the equilibrium properties of biomolecules.
Our results indicates that the free energy landscape can be recovered by 
out-of-equilibrium manipulations, and the collapse of the reconstructed curve
represents an effective criterion to evaluate the reliability of the results.

In conclusion, we believe that our model is an useful tool to investigate the mechanical unfolding of proteins. We plan to further extend our work by studying the single folding and unfolding paths of proteins, by analyzing the unfolding kinetics  of  single substructures, so as to compare the results
with experimental results.

\begin{acknowledgements}
We thank A. Szabo for interesting discussions and  J. Klafter  for his interest in our work.
\end{acknowledgements}

\appendix
\section{Evaluating the free energy landscape 
   from the equilibrium exact solution}\label{FEL}

In this appendix we discuss how the summation in the partition
function (\ref{zeta0}) can be performed exactly, so as to obtain the
free energy landscape as a function of the molecule length
(\ref{effe0}). A key point in our approach is to specify a (finite)
discrete set of values for the length of a molecule. This is of course
not a dangerous assumption, since atomic coordinates in the pdb are
given with a finite resolution and it is therefore fair to round the
distances $\lij$ (measured in $\AA$) to rational numbers with a
finite number of digits. In the applications reported in the present
paper we use the resolution $10^{-3}$ \AA. For the sake of
simplicity, in the following we will adopt a length unit such that the
distances $\lij$ are integer numbers. Notice also that their
absolute value is not greater than $L_{\rm max} = \sum_{i=0}^{N}l_{ii+1}$,
 which corresponds to the length of the molecule in the completely unfolded, fully stretched configuration.

We compute the partition function (\ref{zeta0}) within a recursive
scheme, considering sequences of sub-chains made of peptide bonds from
1 to $n \le N$. For the sub-chain with $n$ bonds we define the
interaction energy
\begin{equation}
  E_{n}(m)=-\sum_{i=1}^{n-1}\sum_{j=i+1}^{n}\epsilon_{ij}\Delta_{ij}\prod_{k=i}^{j}m_{k} 
\end{equation}
and the length 
\begin{equation}
L_{n}(m,\sigma)=\sum_{i=0}^{n}\sum_{j=i+1}^{n+1}l_{ij}\sigma_{ij}S_{ij}(m).
\end{equation}
Notice that $E_{N}(m)-f L_{N}(m,\sigma)$ corresponds to the
Hamiltonian (\ref{hnoi}). We will also need the reduced partition function
\begin{equation}
\Xi_{n}(f)= \sum_{m} \sum_{\sigma} \exp[-\beta E_{n}(m)+\beta f L_{n}(m, \sigma)],
\label{xi}
\end{equation}
where the first summation is over the first $n$ peptide bond variables
and the second one is over the orientations of the corresponding stretches.

Since the length values are integer by choice, we can 
expand $\Xi_{n}(f)$ in powers of $\text{e}^{\beta f}$ as 
\begin{equation}
\Xi_{n}(f)=\sum_{L=-L_{\rm max}}^{+L_{\rm max}}\Ze_{n}(L)\text{e}^{\beta f L}.
\end{equation}
Our goal is thus the evaluation of $\Ze_{N}(L)$, which corresponds to the partition function $Z_0(L)$ (\ref{zeta0}) in the main text.

Using the identity
\begin{equation}
1=1-m_{n}+\sum_{i=1}^{n}(1-m_{i-1})\prod_{k=i}^{n}m_{k},
\end{equation}
one can verify the recursive relations
\begin{equation}
\Xi_{n}(f)=2\sum_{i=1}^{n+1}\cosh(\beta f l_{i-1 n+1})A_{n}^{i}(f),
\end{equation}
and
\begin{equation}
A_{n}^{i}(f) = \left\{\begin{array}{ll}
\exp\pq{\beta\sum_{k=i}^{n}\epsilon_{kn}\Delta_{kn}}A_{n-1}^{i}(f)\; \mbox{
if $i\leq n$;}\\ \Xi_{n-1}(f)\; \mbox{ if $i=n+1$.}
\end{array}\right.
\end{equation}
where, for $i\in\{1,\ldots,n+1\}$,
\begin{eqnarray}
\nonumber A_{n}^{i}(f) & \doteq &
\sum_{m}\sum_{\sigma}(1-m_{i-1})\prod_{k=i}^{n}m_{k}\\
& \cdot & \exp[-\beta E_{n}(m)+\beta f L_{n}(m,\sigma)].
\end{eqnarray}
where the sums over $m$ and $\sigma$ run as in eq.~(\ref{xi}).
In the case $n=0$ we set $\Xi_{0}(f) = 2\cosh(\beta f l_{11})$ and
$A_{0}^{1}(f)=1$.  

Expanding $A_{n}^{i}$ in powers of $\text{e}^{\beta f}$ as
\begin{equation}
A_{n}^i(f)=\sum_{L=-L_{\rm max}}^{+L_{\rm max}}a_{n}^i(L) \text{e}^{\beta f L},
\end{equation}
we see that the coefficients of the above expansion satisfy 
\begin{equation}
a_{n}^{i}(L) = \left\{\begin{array}{ll}
\exp\pq{\beta\sum_{k=i}^{n}\epsilon_{kn}\Delta_{kn}}a_{n-1}^{i}(L)\; \mbox{
if $i\leq n$;}\\ \Ze_{n-1}(L)\; \mbox{ if $i=n+1$,}
\end{array}\right.
\end{equation}
with the initial condition $a_{0}^{1}(L)=\delta(L)$. In addition we have
\begin{equation}
\Ze_n(L)=\sum_{i=1}^{n+1}
\left[a_{n}^{i}(L-l_{i-1 n+1})+a_{n}^{i}(L+l_{i-1 n+1})\right].
\end{equation} 
Notice that the parity with respect to $L$ can be exploited to reduce
the above scheme to positive $L$ values. 

Finally, we want to stress the importance of the integer character of
the microscopic lengths $l_{ij}$. We do not know {\it a priori} the
values that the length of the molecule can assume, so our algorithm
needs to span the whole interval $[0,L_{max}]$, that has thus to be a
finite set. On the other hand, the above scheme can be viewed as a
polynomial algorithm to find the values that $L$ can assume: there are
no configurations $(m,\sigma)$ with length $L$ when $\Ze_{N}(L)=0$.


\begin{thebibliography}{}
\bibitem{ABL} B. Alberts, A. Johnson,  J. Lewis, M. Raff, K. Roberts,  P. Walter, {\it Molecular Biology of the Cell}, (Garland, New York, 2002).
\bibitem{KSGB} M. S. Kellermayer, S. B. Smith, H. L. Granzier C. Bustamante
{\it Science} {\bf 276},{1112} (1997).
\bibitem{rgo}
 M. Carrion-Vasquez {\it et al}, PNAS {\bf 96}, 3694 (1999). 
\bibitem{cv1} M.  Carrion-Vasquez {\it et al}, {\it Nat. Struct. Biol.} {\bf 10}, 738 (2003); H. Dietz, M. Rief PNAS {\bf 101}, 16192 (2004).
\bibitem{DBBR}  H. Dietz, F. Berkemeier, M. Bertz, M. Rief,  PNAS {\bf 103}, 12724 (2006).
\bibitem{Ober1} A. F. Oberhauser, P. K. Hansma, M. Carrion-Vazquez,  J. M. Fernandez, {\it PNAS} {\bf 98}, 468 (2001).
\bibitem{exp_fc1} J.M. Fernandez, H. Li, {\it Science} {\bf 303}, 1674 (2004).
\bibitem{exp_fc2} M. Schlierf, H. Li, J. M. Fernandez PNAS {\bf 101}, 7299 (2004).
\bibitem{NA_exp} C. Danilowicz  {\it et al}  PNAS {\bf 100}, 1694 (2003); J. Liphardt {\it et al}, {\it Science} {\bf 292}, 733 (2001) ; B. Onoa {\it et al} {\it Science} {\bf 299}, 1892 (2003).
\bibitem{thiru} D. K. Klimov, D. Thirumalai {\it Proc. Natl. Acad. Sci. U.S.A} {\bf 97} 7254 (2000).
\bibitem{LKH}  M. S. Li, M. Kouza, C.-K. Hu, {\it  Biophysical Journal} {\bf 92},  1 (2007).
\bibitem{ciep}  P. Szymczak, M. Cieplak,
{\it J. Phys.: Condens. Matter} {\bf 18},  L21 (2006). 
\bibitem{ILT} A. Imparato, S. Luccioli, A. Torcini, e-print  arXiv:0705.3256.
\bibitem{IPZ} A. Imparato, A. Pelizzola, M. Zamparo, {\it Phys. Rev. Lett.} {\bf 98}, 148102 (2007).
\bibitem{WS1} H. Wako, N. Sait\^{o},  {\it J. Phys.~Soc.~Jpn} {\bf 44}, 1931 (1978).
\bibitem{WS2} H. Wako, N. Sait\^{o},  {\it J. Phys.~Soc.~Jpn} {\bf 44}, 1939 (1978).
\bibitem{jarz} C. Jarzynski, {\it Phys. Rev. Lett.} {\bf 78}, 2690 (1997); C. Jarzynski, {\it Phys. Rev. E} {\bf 56}, 5018 (1997); G. E. Crooks,  {\it J. Stat. Phys.} {\bf 90}, 1481 (1998); G. E. Crooks, {\it Phys. Rev. E} {\bf 61}, 2361 (1999).
\bibitem{HumSza} G. Hummer, A. Szabo, {\it Proc. Natl. Acad. Sci. USA} {\bf 98}, 3658 (2001).
\bibitem{alb1}  A. Imparato, L. Peliti,  {\it  J. Stat. Mech.-Theory Exp.}  P03005 (2006).
\bibitem{SFM} M. Sadqi, D. Fushman,  V. Mu\~noz, {\it Nature} (London) {\bf 442}, 317 (2006); F. Huang, S. Sato, T. D. Sharpe, L. Ying,  A. Fersht, {\it Proc. Natl. Acad. Sci. U.S.A.} {\bf 104}, 123 (2007).
\bibitem{dib}  N. Ferguson, T. D. Sharpe, C. M. Johnson, P. J. Schartau, A. R.
Fersht, {\it Nature} (London) {\bf 445}, E14 (2007);
 Z. Zhou and Y. Bai, {\it Nature} (London) {\bf 445}, E16 (2007);
 M. Sadqi, D. Fushman,  V. Mu\~noz, {\it Nature} (London) {\bf 445}, E17 (2007).
\bibitem{ME1} V. Mu\~{n}oz, P.A. Thompson, J. Hofrichter,
W.A. Eaton,  {\it Nature} {\bf 390}, 196 (1997).
\bibitem{ME2} V. Mu\~{n}oz,
E.R. Henry, J. Hofrichter,  W.A. Eaton, {\it Proc. Natl. Acad. Sci. USA}
{\bf 95}, 5872 (1998).
\bibitem{ME3}  V. Mu\~{n}oz, W.A. Eaton, {\it Proc. Natl. Acad. Sci. USA}  {\bf 96}, 11311 (1999).
\bibitem{Amos} A. Flammini,  J.R. Banavar, A. Maritan,
 {\it Europhys. Lett.}  {\bf 58}, 623 (2002).
\bibitem{CCBM} I. Chang, M. Cieplak, J.R. Banavar, A. Maritan A.  {\it Protein Sci.} {\bf 13}, 2446 (2004).
\bibitem{ItohSasai} K. Itoh K.  M. Sasai, {\it Proc. Natl. Acad. Sci. U.S.A} {\bf 101}, 14736 (2004);
K. Itoh, M. Sasai,  {\it Proc. Natl. Acad. Sci. USA}  {\bf 103}, 7298 (2006).
\bibitem{AbeWako} H. Abe, H. Wako,{\it Phys. Rev. E} {\bf 74}, 011913 (2006).
\bibitem{TD1}  V.I. Tokar, H. Dreyss\'e   {\it  Phys. Rev. E}  {\bf 68}, 011601 (2003);
V. I. Tokar, H. Dreyss\'e,  {\it  J. Phys. Cond. Matter }
  {\bf 16}, S2203 (2004); V.I. Tokar,  H. Dreyss\'e, {\it Phys. Rev. E}  {\bf 71}, 031604 (2005).
\bibitem{Ap1}  P. Bruscolini, A. Pelizzola  {\it Phys. Rev. Lett.} {\bf 88}:258101 (2002).
\bibitem{Ap2} A. Pelizzola, {\it  J. Stat. Mech.-Theory Exp.} P11010 (2005).
\bibitem {ZP0} M. Zamparo, A. Pelizzola,  {\it Phys. Rev. Lett.} {\bf 97}, 068106 (2006).
\bibitem {ZP} M. Zamparo, A. Pelizzola, {\it J. Stat. Mech.-Theory Exp.} P12009 (2006).
\bibitem{BPZ1}  P. Bruscolini ,  A. Pelizzola, M. Zamparo, {\it  J. Chem. Phys.} {\bf 126}, 215103 (2007).
\bibitem{BPZ2}  P. Bruscolini ,  A. Pelizzola, M. Zamparo to appear in {\it Phys. Rev. Lett.}.
\bibitem{ev2} E. Evans,  {\it  Annu. Rev. Biophys. Biomol. Struct.}  {\bf 30}, 105 (2001).
\bibitem{Zwa} R. Zwanzig, {\it Nonequilibrium Statistical mechanics}, Oxford University Press, 2001.
\bibitem{denis2}I. Der\'enyi, D. Bartolo,    A. Ajdari, {\it Biophys. J.} {\bf 86}, 1263 (2004).
\bibitem{denis1}D. Bartolo, I. Der\'enyi,  A. Ajdari, Phys. Rev. E {\bf 65}, 051910 (2002).
\bibitem{alb2} A. Imparato, L. Peliti, {\it Eur. Phys. J. B} {\bf 39}, 357 (2004).
\bibitem{ev1}  E. Evans, K. Ritchie {\it Biophys. J.} {\bf 72}, 1541 (1997). 
\bibitem{hs_paper}   O. K. Dudko, G. Hummer, A. Szabo {\it Phys. Rev. Lett.} {\bf 96}, 108101 (2006).
\bibitem{Dudk} O. K.  Dudko {\it et al.}, {\it Proc. Natl. Acad. Sci. U.S.A} {\bf 100}, 11378 (2003).
\bibitem{ferr} A. M. Ferrenberg, R. H.  Swendsen, {\it Phys. Rev. Lett} {\bf 63}, 1195 (1989).
\end{thebibliography}
\end{document}